\documentclass[aps,pra,showpacs,twoside,twocolumn,longbibliography,10pt]{revtex4-1}
\usepackage[colorlinks=true, citecolor=red, urlcolor=blue ]{hyperref}
\usepackage{xcolor}
\usepackage{epsfig,newlfont,amssymb,amsfonts,amsmath,bm,subfigure,palatino,mathtools,amsthm,braket,times,soul,enumitem,color}
\usepackage[normalem]{ulem}
\newcommand{\stkout}[1]{\ifmmode\text{\sout{\ensuremath{#1}}}\else\sout{#1}\fi}
\usepackage[english]{babel}
\usepackage[utf8]{inputenc}
\usepackage{array}
\usepackage{graphics}
\newtheorem{theorem}{Theorem}

\newcommand{\ketbra}[2]{|#1\rangle \langle #2|}
\def\Tr{\text{Tr}}
\newcommand{\kett}[1]{|#1\rangle}
\newcommand{\braa}[1]{\langle #1|}
\begin{document}

\title{
Noncompletely Positive Quantum Maps Enable Efficient Local Energy Extraction in Batteries}

\author{Aparajita Bhattacharyya, Kornikar Sen, Ujjwal Sen}
\affiliation{Harish-Chandra Research Institute, A CI of Homi Bhabha National Institute, Chhatnag Road, Jhunsi, Prayagraj 211 019, India}

\begin{abstract}
Energy extraction from quantum batteries by means of completely positive trace-preserving (CPTP) maps leads to the concept of CPTP-local passive states, which identify  bipartite states from which no energy can be squeezed out by applying any CPTP map to a particular subsystem. We prove, for arbitrary dimension, that if a state is CPTP-local passive with respect to a Hamiltonian, then an arbitrary number of copies of the same state - including an asymptotically large one - is also CPTP-local passive. We show further that energy can be extracted efficiently from  CPTP-local passive states employing NCPTP but still physically realizable maps on the same part of the shared battery on which operation of CPTP maps were useless. Moreover, we provide the maximum extractable energy using local-CPTP operations, and then, we present an explicit class of states and corresponding Hamiltonians, for which the maximum can be outperformed using physical local NCPTP maps. We provide a necessary and sufficient condition and a separate necessary condition for an arbitrary bipartite state  to be unable to supply any energy using non-completely positive trace-preserving (NCPTP)  operations on one party with respect to an arbitrary but fixed Hamiltonian. We build an analogy between the relative status of CPTP and NCPTP operations for energy extraction in quantum batteries, and the association of distillable entanglement with entanglement cost for asymptotic local manipulations of entanglement. The surpassing of the maximum energy extractable by NCPTP maps for CPTP-passive as well as for CPTP non-passive battery states can act as detectors of non-CPTPness of quantum maps.

\end{abstract}

\maketitle

\noindent \emph{\textbf{Introduction.}}-
In recent years, 
research in the miniaturization of devices and in quantum technology~(see e.g.~\cite{Gisin,Peev,Zeilinger,Briegel,Petersen}) has increased significantly, which has led to the investigation of 
quantum batteries~\cite{Alicki,Modi1,Modi2,Santos,Pollock,Sun,Srijon1,Polini,Zhao,Srijon2,Acin,Kornikar,Mir,Frey,Alhambra,manybody1,Tanoy2,Tanoy1,Tanoy3,Kornikar2,kamin,Modi3,Zhang,Ferraro_2018,Crescente,Martin,Giovannetti2,entanglement0,Acin,Gumberidze_2019}.  There have also been attempts to extract maximal energy from these batteries in minimal time, aided by optimal control~\cite{sai,ref1,ref2}. 

There exist states from which no energy can be extracted using unitaries. They are called ``passive states" which depend on the choice of the Hamiltonian~\cite{Lenard,Pusz,Huber,Brown_2016,Sparaciari_2017,Kalu,Silva}.
In the context of open quantum system~\cite{Frey,Alhambra}, shared quantum batteries have been considered, and to extract energy from them, completely positive trace-preserving (CPTP) maps were operated on one fixed part of a bipartite system. 
In this framework also, some states may exist from which no energy can be extracted by operating CPTP maps on the fixed party. These states are identified in Refs.~\cite{Frey} and \cite{Alhambra} and given the names ``strong local passive" and ``CP-local passive" respectively.


The study of open systems beyond the realm of CPTP maps is less explored. In this paper, we explore energy extraction from shared quantum batteries by considering all quantum maps, that include also certain NCPTP maps, in the set of allowed operations on a particular subsystem of the bipartite battery. However, to describe the evolution of physical systems, the NCPTP maps must still be positive, and moreover, not all positive NCPTP maps are quantum mechanically valid ones.

The objective of this work is to look into the hierarchy between physically realizable NCPTP maps and CPTP dynamics in the context of quantum batteries. In this regard, in the first part of our work, we show that energy can be extracted from various two-qubit CPTP-local passive states using NCPTP operations on the same party where application of CPTP operations was unable to extract energy. The successful extraction of energy from such a state can act as a witness for the detection of the non-complete positiveness of these maps. Moreover, we find that whatever be the number of copies, multiple copies of a CPTP-local passive state remain CPTP-local passive with respect to the relevant Hamiltonian. This characteristic of CPTP-local passive states implies that the gap between CPTP and NCPTP local maps,  with respect to CPTP-local passive states, persists in the limit of an asymptotic number of copies.
Further, we show the advantage of NCP maps over CP maps can also be witnessed for general battery states which may not be CP-local passive. 

We draw an analogy between these features and irreversibility in entanglement manipulations as seen in the qualitative and quantitative differences between distillable entanglement and entanglement cost~\cite{HHH,Vidal_Cirac,Vidal_Cirac_pra}. See also Refs.~\cite{acta}, \cite{gisinlink,gisinprl}, and \cite{Horodecki,Lewenstein2, Modi4,Anindita,Plenio,Plenio2}  in this regard. 
Finally we find a necessary and sufficient condition and an independent necessary condition for a bipartite state to be NCPTP-local passive.

Throughout the paper, we use the following nomenclatures: By local CPTP (NCPTP) operations, we mean CPTP (positive trace-preserving) maps acting on a particular subsystem of a bipartite battery while the other part of the battery is kept unchanged. In shorthand notation, we will refer to such local CPTP and NCPTP maps as LCPTP and LNCPTP maps, respectively. Systems that are allowed to have only local CPTP evolution will be specified as CPTP-open systems, whereas systems that can experience NCPTP dynamics are named NCPTP-open systems. 
For a detailed discussion on energy extraction from closed and open quantum batteries, one can go through the Supplementary material (SM).
\vspace{0.2cm}

\noindent 
\emph{\textbf{CPTP-local passivity in the asymptotic limit.-}}
Here we present an interesting property of CPTP-local passive states.
\begin{theorem}
\label{CPTP-local}
If a state, $\rho$, is CPTP-local passive with respect to a fixed Hamiltonian, $H$, then $\rho^{\otimes n}$ will also be CPTP-local passive for any positive integer $n$, with respect to $\sum_n H$.
\end{theorem}

\noindent \emph{Remark.} Here, 
\(\sum_nH\) denotes a sum of Hamiltonians of independent systems having \(H\) as the Hamiltonian for each of them, so that \(\sum_nH = H\otimes I^{\otimes (n-1)} + I \otimes H \otimes I^{\otimes (n-2)} + \ldots + I^{\otimes (n-1)} \otimes H\), where \(I\) denotes the identity operator on the Hilbert spaces of the individual copies.\\

\noindent \textit{Proof.}
Suppose a state, $\rho_{AB}$, which acts on a composite Hilbert space, $\mathcal{H}_{A}\otimes \mathcal{H}_{B}$, is CPTP-local passive with respect to a Hamiltonian, $H_{AB}$. 
Let us consider two copies of $\rho_{AB}$, i.e.,
$\rho_2=\rho_{A_1B_1}\otimes \rho_{A_2B_2}$, and the corresponding Hamiltonian is $H_{A_1B_1A_2B_2}=H_{A_1B_1}\otimes I_{A_2B_2}+I_{A_1B_1}\otimes H_{A_2B_2}$. 
Let us introduce two more Hilbert spaces, $\mathcal{H}_{A_1'}$ and $\mathcal{H}_{A_2'}$, which are copies of $\mathcal{H}_{A_1}$ and $\mathcal{H}_{A_2}$ respectively, and define an operator $C_{A_1A_2A_1'A_2'}=\text{Tr}_{B_1B_2} \left[\left(\rho_{2}^{\text{T}_{A_1}\text{T}_{A_2}} \otimes {I}_{A_1'A_2'}\right) \left({I}_{A_1A_2} \otimes H_{B_1B_2A_1'A_2'}\right)\right]$. Though $\mathcal{H}_{A_1}$ and $\mathcal{H}_{A_2}$ have the same dimension, let us denote the dimension of each of these Hilbert spaces as $d_{A_1}$ and $d_{A_2}$, respectively.
It is easy to check that
$C_{A_1A_2A'_1A'_2}=C_{A_1A_1'}\otimes M_{A_2}\otimes I_{A_2'} + M_{A_1}\otimes I_{A_1'} \otimes C_{A_2A_2'}, \quad$
where 
$M_{A_i}=\text{Tr}_{B_i}\left(\rho_{A_iB_i}^{T_{A_i}}\right)=\left[\text{Tr}_{B_i}(\rho_{A_iB_i}) \right]^T\ge 0$ and
$C_{A_iA_i'}=\text{Tr}_{B_i} \left[\left(\rho_{A_iB_i}^{\text{T}_{A_i}} \otimes {I}_{A_i}\right)\left({I}_{A_i'}\otimes H_{B_iA_i'}\right)\right].$
Since both $\rho_{A_1B_1}$ and $\rho_{A_2B_2}$ are CPTP-local passive, from the condition of CPTP-local passivity 
we have
\begin{equation}
\label{eq}
    C_{A_iA_i'} - \text{Tr}_{A_i'}\left(d_{A_i}^2 \ketbra{\phi^+}{\phi^+} C_{A_iA_i'} \right) \otimes \mathcal{I}_{A_i'} \geq 0~\text{for }i=1,2.
\end{equation}
We want to check if the condition of CPTP-local passivity also holds for $\rho_2$, i.e., if the inequality
still remains satisfied if we replace $A$ and $A'$ by $A_1A_2$ and $A_1'A_2'$, respectively. By doing so on the left-hand side of the CPTP-local passivity condition,
we get
\begin{widetext}
\begin{eqnarray}
\label{cp_2copy2}
 C_{A_1A_2A'_1A'_2} - \text{Tr}_{A'_1A'_2}\left (d_{A_1A_2}^2 \ketbra{\phi^+}{\phi^+} C_{A_1A_2A'_1A'_2} \right ) \otimes \mathcal{I}_{A'_1A'_2}
   = (M_{A_1}\otimes I_{A_1'}) \otimes Q_{A_2A_2'} + 
     Q_{A_1A_1'} \otimes (M_{A_2}\otimes I_{A_2'}),\label{eq2}
\end{eqnarray}
\end{widetext}
where $Q_{A_iA_i'}=C_{A_iA_i'} - \text{Tr}_{A_i'}\left (d_{A_i}^2 \ketbra{\phi^+}{\phi^+} C_{A_iA_i'} \right ) \otimes \mathcal{I}_{A_i'}\geq 0 $ [see Eq.~\eqref{eq}].
Here we have used the relation that $\text{Tr}_{A_1'}\left [(\ket{\phi^+}\bra{\phi^+}\right)_{A_1A_2A'_1A'_2} ] = \frac{1}{d_{A_1A_2}}(\ket{i}\bra{i})_{A_1} \otimes (\ket{\phi^+}\bra{\phi^+})_{A_2A_2'}.$
The above relation also holds if we swap $A_1$ ($A_1'$) with $A_2$ ($A_2'$). Since $Q_{A_iA_i'},~M_{A_i}\geq 0$, we can say the entire operator   presented on the right-hand side or left-hand side in Eq.~\eqref{eq2} is positive. This implies $C_{A_1,A_2,A_1',A_2'}$ also satisfies the condition of CPTP-local passivity.
Therefore, if we take two copies of CPTP-local passive states, $\rho_{CP}$, then $\rho_{CP}^{\otimes 2}=\rho_{CP} \otimes \rho_{CP}$ is also a CPTP-local passive state. Since this proof is valid for arbitrary dimension of $\mathcal{H}_A, \mathcal{H}_B$ and $\mathcal{H}_{A'}$, following similar argument 
the CPTP-local passivity of $\rho_{CP}^{\otimes 2^n}$, for arbitrary integer, $n$, can  be shown by induction. 
Further, if $2^n$ copies of a state is CPTP-local passive, it is obvious that any $x<2^n$ copies of the same state will also be locally passive for any positive integer, $x$. Hence also in the asymptotic limit, i.e., for $n\rightarrow \infty$, the state would turn out to be CPTP-local passive.\hfill $\blacksquare$





\vspace{0.2cm}

\noindent 
\emph{\textbf{Comparison between CPTP-open and NCPTP-open batteries.-}} Henceforth, in this paper, by NCPTP maps, we will mean only those which are physically realizable, i.e., valid quantum maps. The maximum amount of energy that can be extracted from a battery state, $\rho_{AB}$, using local NCPTP maps can be written as
\begin{eqnarray}
\label{brisTi}
  \Delta W_{\max}^{NCP}(\rho_{AB},H_{AB})=  \text{Tr}\left[\rho_{AB} H_{AB} \right] \nonumber\\-\min_{\Lambda_A} \text{Tr}\big[\Lambda_A\otimes I_B(\rho_{AB}) H_{AB} \big]. 
\end{eqnarray}
Here the minimization is performed over all NCPTP-maps, $\Lambda_A$. 
Before considering general NCPTP maps, $\Lambda_A$, let us first focus on energy extraction using a particular type of LNCPTP evolution. 
To form the LNCPTP dynamics, consider a tripartite system consisting of a bipartite battery, $AB$, and an environment $E$. Let the tripartite initial state, $\rho_{EAB}$, be pure and entangled within the bipartition $E:AB$. The battery state, $\rho_{AB}$, can be obtained by tracing out $E$, i.e., $\rho_{AB}=\text{Tr}_E(\rho_{EAB})$. We want to evaluate the maximum extractable energy from $\rho_{AB}$ for a given Hamiltonian $H_{AB}$, using NCPTP operations on the system $A$. The generation of such an LNCPTP evolution can be ascertained by a global unitary evolution of the environment, $E$, and the system, $A$, followed by tracing out the environment. Therefore, the maximum extractable energy from $\rho_{AB}$, with respect to Hamiltonian $H_{AB}$, using this kind of LNCPTP map is 
\begin{eqnarray}
\label{megh}
   && \Delta W_{p}^{NCP} (\rho_{AB},H_{AB})=  \text{Tr}\left[\rho_{AB}  H_{AB} \right] \nonumber \\ 
   &-& \min_{U_{EA}} \text{Tr}\big[\text{Tr}_E(U_{EA}\otimes {I}_B\rho_{EAB}U_{EA}^{\dagger}\otimes {I}_B) H_{AB} \big],\label{3eq2}
\end{eqnarray}
where $U_{EA}$ are global unitaries acting on $E$ and $A$. Here we have introduced a \emph{particular type} of LNCPTP channel, assuming that $\rho_{AB}$ is part of a tripartite state, $\rho_{EAB}$, which as a whole is pure. The subscript, $p$, used on the left-hand side, helps to keep this fact in mind. The dynamics will be different if $\rho_{EAB}$ is mixed. In such cases, depending on the mixedness, and the dimension of the attached environment, $E$, many other LNCPTP channels can be applied to $\rho_{AB}$ but application of the channels discussed above considering $\rho_{EAB}$ pure may not be implementable.
Note that in effect we are not assuming here a part of the assumption referred to as the church of the higher Hilbert space, viz. 
the existence of an entangled pure state in a larger Hilbert space for every mixed quantum state.
Since 
the expression of $\Delta W_{\max}^{NCP}(\rho_{AB},H_{AB})$
involves maximization over all LNCPTP maps, $\Delta W^{NCP}_{\max}(\rho_{AB},H_{AB})\geq \Delta W^{NCP}_p(\rho_{AB},H_{AB})$ for all bipartite states, $\rho_{AB}$, and Hamiltonian, $H_{AB}$. 

Such an LNCPTP map can be implemented by first generating genuine three-party entangled states whose two-party reduced densities are entangled, such as in $\rho_{EAB}$, and then applying a nonlocal unitary operator on the parties, $E$ and $A$. The $W$ state~\cite{W} or cluster states~\cite{cluster,cluster2} are suitable three-party entangled states, of which the experimental generation of $W$-type states is proposed in optical systems~\cite{WW1,WW2,WW3}, trapped ions~\cite{tr_ion}, atomic ensembles~\cite{at_ens} and integrated nanophotonic circuits~\cite{W1}. 
Cluster states are highly entangled quantum states which can be prepared in spin qubit lattices by interacting them with Ising type Hamiltonian~\cite{cluster}, or 
in photonic systems~\cite{cl2,cl3,cl4}. 
Whereas, non-local unitaries have been experimentally implemented in optical systems~\cite{U1}, solid state systems~\cite{U2}, molecular electron spin qubits~\cite{U3} etc.

In particular, three-mode $W$-type entangled coherent states (ECS)~\cite{ecs,ecs_review} can be produced experimentally in photonic systems~\cite{WW1}. The prepared three-mode ECS can then be fed as input to an optical multiport, generating a nonlocal unitary~\cite{U1} acting on two of the modes. The environmental mode can ultimately be discarded from the final state obtained. 
This can lead to an 
implementation of an LNCPTP map on a two-mode optical system.


The main objective of this paper is to compare $\Delta W^{NCP}_{\max}(\rho_{AB},H_{AB})$ with $\Delta W^{CP}_{\max}(\rho_{AB},H_{AB})$, where $\Delta W^{CP}_{\max}(\rho_{AB},H_{AB})$ is defined in the same way as $W^{NCP}_{\max}(\rho_{AB},H_{AB})$ [see Eq.~\eqref{brisTi}] with the only difference being that the minimization involved in the second term is performed over all CPTP-maps acting on subsystem $A$.

\begin{figure}
\centering
\includegraphics[width=8cm]{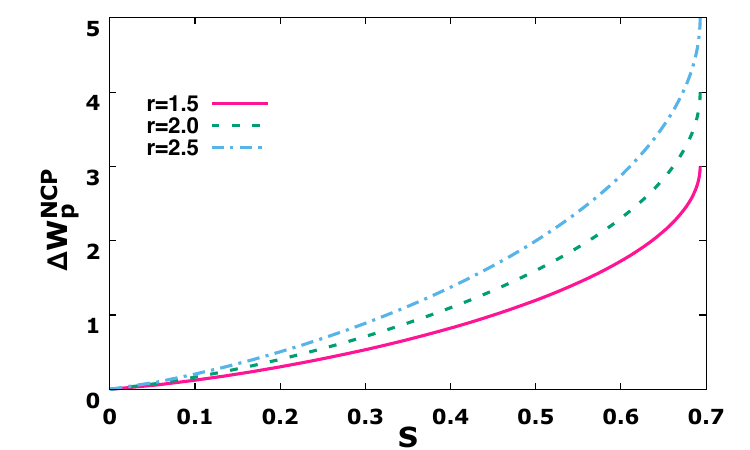}%
\caption{Maximum extractable energy from CPTP-local passive states using LNCPTP operations. 
The behavior of $\Delta W_p^{NCP}$ is shown for various values of $r$, i.e. the interaction strength, corresponding to the numerical values $p=q=e/2$. The horizontal and vertical axes are in units of ebits and energy, $e$, respectively.}
\label{ncp}
\end{figure}

\vspace{0.2cm}

\noindent 
\emph{\uline{Extraction of energy from CPTP-local passive states.}}-
Let us begin by considering the two-Bell mixture battery state, given by
   $ \widetilde{\rho}_{AB}=p_1 \ketbra{\psi^+}{\psi^+} + (1-p_1) \ketbra{\psi^-}{\psi^-}$,
where $\kett{\psi^\pm}$ are maximally entangled states:   $\kett{\psi^+}=(\kett{01}+\kett{10})/\sqrt{2}$ and $\kett{\psi^-}=(\kett{01}-\kett{10})/\sqrt{2}$. Let us also fix the Hamiltonian to be
   $ \widetilde{H}_{AB}=p(\sigma^z \otimes \mathcal{I}) + q(\mathcal{I} \otimes \sigma^z) + r(\sigma^x \otimes \sigma^x + \sigma^y \otimes \sigma^y)$ .
Here $\sigma^x, \sigma^y$ and $\sigma^z$ are the Pauli matrices, and $r$ is the strength of the interaction between $A$ and $B$. For all the numerical calculations, we take $p=q=e/2$, where $e$ has the unit of energy. The state, $\widetilde{\rho}_{AB}$, 
is CPTP-local passive with respect to the Hamiltonian, $\widetilde{H}_{AB}$, 
for $p_1 \le 1/2$ and $r \ge p/(1-2p_1)$~\cite{Alhambra}. 
We now consider the system, $AB$, to be NCPTP-open and examine if energy extraction is possible through LNCPTP operations.
The details on the construction of the LNCPTP map are given in the SM.



Let us consider the set of  $\widetilde{\rho}_{AB}$ which obeys $p_1 \le 1/2$ and $r \ge p/(1-2p_1)$. Since these states are CPTP-local passive, $\Delta W^{CP}_{\max}=0$. 
The behavior of 
$\Delta W^{NCP}_{p}(\widetilde{\rho}_{AB},\widetilde{H}_{AB})$, is depicted with respect to the entanglement, $S$, of $\widetilde{\rho}_{EAB}$, 
the system-environment joint state, in the bipartition $E:AB$, for different values of interaction strength, $r$, in Fig.~\ref{ncp}.  We measure $S$ using entanglement entropy. We obtain that $\Delta W^{NCP}_{p}$ is zero at $S=0$. This is evident as for $S=0$, $\widetilde{\rho}_{EAB}$
is a pure product state in the bipartition $E:AB$ 
and therefore the generated map is still completely positive. 
Further, we realize that with an increase in $S$, extractable energy using the LNCPTP evolution also increases. Hence, we can conclude that energy extraction using LNCPTP dynamics is potentially useful and advantageous over energy extraction using LCPTP processes.
The optimization is performed over all two-qubit unitaries (see~\cite{Kraus}), which includes nonlocal ones, and the maximal extracted energy corresponds to the optimal unitary. Such an unitary can be, for example, generated in phtononic systems using the mechanism given in~\cite{U1}.
We conclude that a state that seemed to be ``passive" under LCPTP maps is no longer ``passive" if more general positive maps, the physical LNCPTP maps, are allowed. Note that by ``passive," we mean that the state is unable to provide any energy through the application of the corresponding map under consideration. Moreover, we observe that the maximum energy extracted using the LNCPTP operation, $\Delta W_p^{NCP}$, increases with increasing strength of the interaction, $r$. 
We have proved in Theorem 1 that CPTP-local passive state remains so when an asymptotically large number of copies of the state is considered. Therefore we can conclude that the gap in the energy extraction using LCPTP and LNCPTP maps will remain intact or be more even if we move towards the asymptotic limit, at least in the case of CPTP-local passive states.

\vspace{0.2cm}

\noindent 
\emph{\textbf{Maximum extractable work using LCPTP operations: Surpassing the CPTP-maximum using LNCPTP operations.-}}
The maximum extractable work using LCPTP operations is given by 
\begin{equation}
\label{max}
\Delta W^{CP}_{max}=\text{Tr}(H_{AB}\rho_{AB})-c_{min},
\end{equation}
where $c_{min}$ is the minimum eigenvalue of $C_{AA'}$. The maximum is valid for states and Hamiltonians for which $C_{AA'}$ has a non-degenerate eigenspectrum. The proof of Eq.~\eqref{max} is given in  the SM.

We find that 
the extractable energy using LNCPTP operations exceeds the CPTP-maximum for a class of states, $\widetilde{\rho}_{AB}$, and Hamiltonian, $\widetilde{H}_{AB}$, under the condition
$
    \frac{3}{8} |r||x| > |r|+(p+q)/2,$
where $r<0$. Here $x=\cos(\alpha_2-\beta_2-\gamma_2)+2\cos(\beta_2+\gamma_2)+\cos(\alpha_2+\beta_2+\gamma_2)-4$, where $\alpha_2,\beta_2$ and $\gamma_2$ are parameters of a unitary, $\mathbb{U}_{2}$, using which, energy is being extracted. The details of the derivation are given in the SM.
\vspace{0.2cm}

\begin{figure}
\centering
\includegraphics[width=8cm]{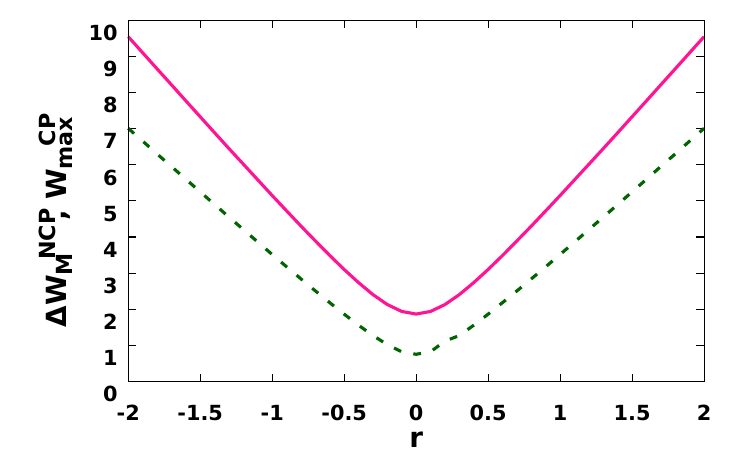}%
\caption{Comparison between LCPTP and LNCPTP operations as  functions of interaction strength, $r$. The numerical values of $p$ and $q$ are the same as in Fig. \ref{ncp}. 
All quantities plotted have the units of energy, $e$.}
\label{cp_ncp_gen}
\end{figure}

\noindent 
\emph{\textbf{Energy extraction from CPTP non-passive batteries using LNCPTP maps.}}-
Here we
consider the same Hamiltonian, $\widetilde{H}_{AB}$, and find the rank-two state, $\Bar{\rho}_{AB}^r$, which provides the maximum amount, among all rank-two states, of extractable energy when LNCPTP operations are applied. A 
discussion on its motivation is provided in the SM.
We denote the amount of extractable energy from $\Bar{\rho}_{AB}^r$ using LNCPTP operations by $\Delta \Bar{W}_M^{NCP}(r)$ which is equal to $\Delta \Bar{W}_M^{NCP}(r)=\max_{\Bar{\rho}_{AB}^r}\Delta W_{p}^{NCP}(\Bar{\rho}_{AB}^r,\widetilde{H}_{AB})$, where $r$ is the coupling between $A$ and $B$, and the maximization is done over all rank two states, $\Bar{\rho}_{AB}^r$. $\Delta \Bar{W}_{M}^{NCP}(r)$ is being written as a function of $r$ because it of course depends on the Hamiltonian, $\widetilde{H}_{AB}$, but for all the numerical calculations, we fix $p$ and $q$ to $\epsilon/2$ and only vary $r$. The dependency of $\Delta \Bar{W}_{M}^{NCP}(r)$ on $r$, 
is illustrated in Fig.~\ref{cp_ncp_gen} 
using a pink line.
The details of the construction of the LNCPTP map are given in the SM. We also calculate the maximum amount of energy, $\Delta 
 {W}_{\max}^{CP}(\Bar{\rho}_{AB}^r,\widetilde{H}_{AB})$, that can be extracted using LCPTP operations, from the same state, $\Bar{\rho}_{AB}^r$ and present the nature of $\Delta W_{\max}^{CP}(\Bar{\rho}_{AB}^r,\widetilde{H}_{AB})$  in the same figure, Fig.~\ref{cp_ncp_gen}, using the green dashed line, with respect to the corresponding value of $r$ for which $\Bar{\rho}_{AB}^r$ has been obtained.
It can be noticed from the figure that $\Delta {W}_{\max}^{CP}(\Bar{\rho}_{AB}^r,\widetilde{H}_{AB})$ is always smaller than $\Delta \Bar{W}_{M}^{NCP}(r)$ which is equal to $\Delta {W}_{p}^{NCP}(\Bar{\rho}_{AB}^r,\widetilde{H}_{AB})$ for all the considered values of $r$.

It should be noted that if instead of rank-two states, rank-one states were considered in place of $\Bar{\rho}_{AB}$ or $\widetilde{\rho}_{AB}$, the energy extraction operation would reduce to a CPTP map. To get the flavor of NCPTP maps, one must assure that the rank of the battery state is higher than unit. For numerical simplicity, we have restricted ourselves to rank-two states. We expect that considering higher rank states would not affect the results qualitatively. 

\vspace{0.2cm}

\noindent 
\emph{\textbf{Detecting NCPTP maps.-}} 
One can notice that Eq.~\eqref{max} can be utilized to formulate a witness for the detection of the non-complete positivity of the governing dynamics.
That is, to check if an unknown map is CP or not, we can take a particular state and try to extract energy from it by operating the unknown map on a particular subsystem of the state.
If 
the amount of energy that can be extracted by operating the map surpasses $\Delta W^{CP}_{\max}$, then the map is certainly non-CPTP. If the energy extraction is less than that, then no inference can be drawn 
about the non-CPTP nature of the unknown map. A particularly interesting case is of a nonzero NCPTP energy extraction from CPTP-passive batteries.

\vspace{0.2cm}

\noindent 
\emph{\textbf{Analogy with asymptotic manipulations of entanglement.-}}
On a similar footing with the relation between distillable entanglement, $E_D$, and entanglement cost, $E_C$ (see Fig.2-(a) of SM), we find that when $\Delta {W}_{\max}^{CP}=0$, i.e., for CPTP-local passive states, $\Delta {W}_{p}^{NCP}$ can be positive. Moreover, even when $\Delta {W}_{\max}^{CP}>0$, we have seen that $\Delta {W}_{p}^{NCP}$ can also be positive and strictly greater than $\Delta {W}_{\max}^{CP}$. 
This fact is depicted in Fig.2-(b) of SM.
We have considered only those LNCPTP maps that can be generated by applying unitaries on the pure state $\kett{\Bar{\Psi}}_{EAB}$. The situation can be generalized by considering mixed states of the entire system, $EAB$, which may further enhance the extractable energy. It can therefore be potentially inferred that $\Delta {W}_{\max}^{CP}$ and $\Delta {W}_{\max}^{NCP}$ exhibit a relation similar to that between 
$E_D$ 
and 
$E_C$ respectively.
This is merely a coincidence as of now. We however believe that there might be connections between the two.
This is because the target state in analyzing entanglement cost is a mixed state in cases when there is a finite gap between entanglement cost and distillable entanglement. This is similar to what happens in NCPTP energy extraction - the state of $AB$ is necessarily mixed, as otherwise, the operation on $A$ cannot be a (physical) NCPTP map. 
The target state in entanglement distillation is a pure state (asymptotically), and in parallel, if $AB$ is in a pure state, the operation on $A$ is necessarily CPTP. 

%

\vspace{0.2cm}

\noindent  
\emph{\textbf{Necessary and sufficient condition for NCPTP-local passivity.-}} Next we provide a necessary and sufficient condition for a state to be NCPTP-local passive for an arbitrary but fixed Hamiltonian, $H_{AB}$.
The pair, $\{H_{AB}, \rho_{AB}\}$, is NCPTP-local passive, with respect to subsystem $A$ where $\rho_{AB}$ is attached to an environment, $E$, and the complete state of the composite system, $ABE$, is $\rho_{ABE}$, if and only if 
$C-C'\geq 0$,
where $ \braket{\alpha\beta|C'|\alpha' \beta'}=\delta_{\alpha \alpha'} \sum_i \braket{ii|C|\beta' \beta}$. Here  $\ket{i} \in \mathcal{H}_{E_i} \otimes \mathcal{H}_{A_i}$, $\ket{\beta^{(')}} \in \mathcal{H}_{E_{1(2)}} \otimes \mathcal{H}_{A_{1(2)}}$, $C\equiv C_{E_1 A_1 E_2 A_2}=\bra{\phi}\left(\mathcal{I}_{E_1}\otimes H_{A_1 B_1}\otimes\rho_{E_2 A_2 B_2}^T \right) \ket{\phi}_{B_1 B_2}$, and $\ket{\phi}_{B_1B_2}$ is the unnormalized maximally entangled state, given by $\ket{\phi}_{B_1B_2}=\sum_j \ket{jj}$. $H_{A_1B_1}$ and $\rho_{E_2A_2B_2}$ are copies of
$H_{AB}$ and $\rho_{ABE}$, respectively, which act on
 the Hilbert spaces $\mathcal{H}_{A_1}\otimes\mathcal{H}_{B_1}$ and $\mathcal{H}_{E_2}\otimes\mathcal{H}_{A_2}\otimes\mathcal{H}_{B_2}$ respectively, where $\mathcal{H}_{E_i}$, $\mathcal{H}_{A_i}$ and $\mathcal{H}_{B_i}$ denote the Hilbert spaces of, respectively, the environment and subsystems $A$ and $B$ for $i=1$ and $2$. 
The proof of this statement is given in SM (see~\cite{Magnus,schur} for  references). Additionally, 
we provide an independent necessary condition for NCPTP-local passivity using a different approach in the SM.

\vspace{0.2cm}

\noindent 
\emph{\textbf{Conclusion.-}} We examined the contrast between CPTP-open and NCPTP-open quantum batteries. The actions of CPTP and NCPTP  maps have been restricted to only a subsystem. 
CPTP-local passive states of a bipartite quantum battery, described by a fixed Hamiltonian, are those states from which no energy can be extracted by applying CPTP maps to a particular part of the battery.
One of our main goals in this paper is 
to go beyond CPTP maps and compare the efficacy of physical NCPTP maps with that of CPTP maps in the context of energy extraction from quantum batteries. In this regard, we first proved that CPTP-local passive states remain CPTP-local passive even in the asymptotic limit and then showed that energy can be extracted from these CPTP-local passive states using LNCPTP  maps.
The extracted energy using NCPTP  maps from CPTP-local passive states can be considered as a witness for the detection of the NCPTP  nature of the applied map. 
 Moreover, we maximized the energy extractable from a quantum battery using CPTP-local operations, and found even if this  maximum 
 is positive, the extractable energy using LNCPTP  operations can exceed that positive value. These features are 
analogous to relative values of distillable entanglement and entanglement cost, where there exist shared quantum states for which the former vanishes while the latter is non-zero, as well as quantum states for which the former is non-zero and yet the latter is even larger.
Finally, we presented a necessary and sufficient condition and an independent necessary condition for an arbitrary state to be NCPTP-local passive, which depends on the form of the Hamiltonian.

\vspace{0.2cm}

 \noindent 
\emph{\textbf{Acknowledgements.-}}We acknowledge computations performed using Armadillo~\cite{article,try}
   on the cluster computing facility of the Harish-Chandra Research Institute, India.
   The research of KS was supported in part by the INFOSYS scholarship.
   We acknowledge partial support from the Department of Science and Technology, Government of India through the QuEST grant (grant number DST/ICPS/QUST/Theme-3/2019/120).



\begin{center}
    \textbf{Supplementary Material for\\Non-completely positive quantum maps enable efficient local energy extraction in batteries}
\end{center}

\maketitle
\section{Introduction}
The storage of energy in a system and the extraction of maximum amount of the stored energy constitute the fundamental aspects of energy-storing devices such as batteries.
Ordinary batteries can store chemical energy and transform it into electrical energy whenever necessary. For such macroscopic systems, energy exchange between the system and its environment is governed by the laws of thermodynamics. In recent years, however, research in the miniaturization of technological devices has increased significantly, which has led to the investigation of small-sized batteries~\cite{Alicki,Modi1,Modi2,Santos,Pollock,Sun,Srijon1,Polini,Zhao,Srijon2}, in which quantum characteristics like entanglement~\cite{Horodecki,Lewenstein2}, quantum discord~\cite{Modi4,Anindita},  quantum coherence~\cite{Plenio,Plenio2} can be present. 
These quantum batteries are expected to be relevant to modern-day technologies like communication~\cite{Gisin,Peev,Zeilinger}, computation~\cite{Briegel}, optimal control~\cite{Petersen}, and so on.

The concept of quantum batteries was, as far as we know, first formalized by Alicki and Fannes in 2013~\cite{Alicki}. Significant advancements have since been made in the research on closed~\cite{Acin,Kornikar,Mir} and open~\cite{Frey,Alhambra,manybody1,Tanoy2,Tanoy1,Tanoy3,Kornikar2,kamin,Srijon2} quantum batteries, including many body quantum batteries, where several models like short and long-range XXZ quantum spin chains~\cite{Modi3}, spins in cavities governed by the Dicke interaction~\cite{Zhang,Ferraro_2018,Crescente,Martin,Giovannetti2}, the ordered and disordered XYZ model~\cite{Srijon1}, etc. have been explored.
The role of quantum resources like entanglement and quantum coherence in the performance of quantum batteries has also been delved into~\cite{entanglement0}.
For instance, in Ref.~\cite{Acin}, the authors have shown that the generation of entanglement is not necessary for maximal energy extraction from batteries. 
Further studies have showed that it is possible to increase both the energy and quantum coherence of a system by utilizing partial coherence from an environment, and making a measurement on the system~\cite{Gumberidze_2019}.

The internal energy of a system can be thought as comprising of two parts: ``ergotropy", i.e., the maximum extractable energy from a system using unitary operations, and the other part, termed ``passive energy", which is the amount of energy that cannot be extracted using unitaries. There exist states that only consist of passive energy; therefore, it is impossible to extract energy from these states using unitary evolution. They are called ``passive states"~\cite{Lenard,Pusz,Huber,Brown_2016,Sparaciari_2017,Kalu}. The set of passive states depends on the choice of the Hamiltonian.

Passivity and energy extraction are extensively studied in the context of quantum batteries. One of the initial works in the direction of the extraction of energy from open quantum batteries was put forward by Frey \emph{et al}.~\cite{Frey}, and followed by Alhambra \emph{et al}.~\cite{Alhambra}. In these works, shared quantum batteries were considered, and to extract energy from them, completely positive trace-preserving (CPTP) maps were operated on one fixed part of a bipartite system. 
In this framework also, some states may exist from which no energy can be extracted by operating CPTP maps on the fixed party. These states are identified in Refs.~\cite{Frey} and \cite{Alhambra} and given the names ``strong local passive" and ``CP-local passive" respectively.


The maps describing the evolution of quantum mechanical systems are considered to be linear, which transforms a set of density matrices into the same or another set of density matrices. The properties of density matrices further restrict the mapping to being trace-preserving, hermiticity-preserving, and positive. Complete positivity is also often imposed.
Therefore, CPTP maps are a subclass of all possible linear maps that can describe the evolution of physical systems. Despite the fact that CPTP maps do not describe all open system quantum dynamics, the study of open systems beyond the realm of CPTP maps is less explored. In this paper, instead of restricting ourselves to CPTP maps, we explore energy extraction from shared quantum batteries by considering all quantum maps, that include also certain non-completely positive trace-preserving (NCPTP) maps, in the set of allowed operations on a particular subsystem of the bipartite battery. However, to describe the evolution of physical systems, the NCPTP maps must still be positive, and moreover, not all positive NCPTP maps are quantum mechanically valid ones.

The objective of this work is to look into the hierarchy between physically realizable NCPTP maps and CPTP dynamics in the context of quantum batteries. In this regard, in the first part of our work, we show that energy can be extracted from various two-qubit CPTP-local passive states using NCPTP operations on the same party where application of CPTP operations was unable to extract energy. The successful extraction of energy from such a state can act as a witness for the detection of the non-complete positiveness of these maps. Moreover, we find that whatever be the number of copies, multiple copies of a CPTP-local passive state remain CPTP-local passive with respect to the relevant Hamiltonian. This characteristics of CPTP-local passive states implies the gap between CP and NCP local maps,  with respect to CPTP-local passive states, persists in the limit of asymptotic copies. 

We proceed further to investigate if the advantage of NCP maps over CP maps can also be witnessed for general battery states which may not be CP-local passive. In this regard, we first provide the maximum extractable energy by performing local-CPTP operations on one party of a shared battery state and then present a class of two-qubit states and Hamiltonians, which can overcome this maximum if operations in hand are local NCP maps and they are operated on the same party.

We draw an analogy between these features and irreversibility in entanglement manipulations as seen in the qualitative and quantitative differences between distillable entanglement and entanglement cost~\cite{HHH,Vidal_Cirac,Vidal_Cirac_pra}. Se also Refs.~\cite{acta} and \cite{gisinprl}  in this regard. Finally, we also present a necessary condition for a bipartite state to be NCPTP-local passive, i.e., for a state that is unable to provide any energy through any NCPTP operation on one particular party, in terms of the Hamiltonian that describes the energy of the shared quantum battery under consideration. By restricting ourselves to two-qubit rank-two batteries, we have added some more necessary conditions to verify NCPTP-local passive states.

%

\section{Energy extraction from closed and CPTP-open quantum batteries}
\label{sec:2}
Before moving to open quantum batteries, let us begin by briefly recapitulating energy extraction from quantum batteries under closed system dynamics. Typically, a quantum battery is defined in terms of a quantum mechanical system characterized by a state, $\rho$, and a Hamiltonian, $H$. 
A fraction of the stored energy of the battery can be extracted by 
incorporating an extra term in the Hamiltonian of the system, e.g., by
applying a field for a certain duration of time, say $\tau$. The applied field, in general, will cause the battery state to change and transform to a different state, i.e., $U(\lambda,\tau)\rho U(\lambda,\tau)^{\dagger}$, where $U$ represents the unitary operator that expresses the evolution of the system in the presence of the field. Here $\lambda$ denotes the congregate of parameters of the unitary operator. The amount of energy that can be extracted from the battery in this process is 
\begin{equation}
\label{W}
    \Delta W^{U}=\text{Tr}(\rho H) - \text{Tr}\left[U(\lambda,\tau)\rho U(\lambda,\tau)^{\dagger} H\right].
\end{equation}
The prime objective is to extract as much energy as possible from the battery. The maximum extractable energy from a quantum battery by unitary operations is given by
\begin{eqnarray}
\label{erg}
   \Delta W_{\max}^{U} &=& \text{Tr}(\rho H) - \min_U \text{Tr}\left[U(\lambda,\tau)\rho U(\lambda,\tau)^{\dagger} H\right], \nonumber \\
    &=& \text{Tr}(\rho H) -  \text{Tr}(\sigma H).    
\end{eqnarray}
$\Delta W_{\max}^{U}$ is referred to as the ergotropy of the battery, defined by the pair $\{\rho,H\}$, for unitary operations. It is an important figure of merit for a quantum battery. 
The first term on the right-hand side of Eq.~\eqref{erg} represents the total energy of the initial state, $\rho$, corresponding to the given Hamiltonian, $H$. The second term determines the amount of energy remaining in the system after maximum energy extraction from $\rho$ via unitary operations. So $\text{Tr}(\sigma H)$ is the amount of ``forbidden" energy which cannot be extracted from $\rho$, for the given $H$ and would be left in the system after extracting all the unitarily accessible energy from $\rho$.
$\sigma$ is called the passive state. Formally, passive states are states from which energy extraction is not possible using unitary operations. Passive states are crucial elements in the framework of quantum batteries or, more generally, in quantum thermodynamics~\cite{Silva,Huber,Brown_2016,Sparaciari_2017,Kalu}. The necessary and sufficient conditions for a state, $\sigma$, to be passive are that it should commute with its Hamiltonian, $H$, and if for a particular order of the eigenvectors the corresponding eigenvalues of $H$, $\{\epsilon_i\}_i$, satisfy $\epsilon_i>\epsilon_j$, then the eigenvalues of the passive state, $\{\lambda\}_i$, should satisfy $\lambda_i\leq \lambda_j$ for all $i$ and $j$~\cite{Lenard,Pusz}.

We have discussed energy extraction from systems using unitary operations. But in reality, there exist more general physical operations that may not be described by unitary operations but can be formulated using CPTP maps. Before going into the details of energy extraction using CPTP maps, let us briefly recall the origin of these maps. Consider a system, $S$, and an environment, $E$, described by states, $\rho_S$ and $\rho_E$, which act on the two Hilbert spaces, $\mathcal{H}_S$ and $\mathcal{H}_E$, respectively. If it is possible to bring the two systems together, one can operate a joint unitary operation, $U_{SE}$ on the composite state, $\rho_S\otimes\rho_E$. 
Note the tacit assumption that the joint state is a product.
After operating $U_{SE}$, if we ignore the environment $E$, the final state of the system, $S$, can be expressed as
\begin{eqnarray}
    \rho'_S=\mathcal{E}(\rho_S)=\Tr_E\left(U_{SE}\rho_S\otimes\rho_E U_{SE}^\dagger\right)=\sum_\mu K_\mu \rho_S K_\mu^\dagger,\label{3eq1}
\end{eqnarray}
where $\{K_\mu\}_\mu$ is a set of operators, known as Kraus operators, satisfying $\sum_\mu K_\mu^\dagger K_\mu=I_S$ (completeness relation). Here $I_S$ denotes the identity operator acting on $\mathcal{H}_S$. From the completeness relation, it can be easily seen that $\Tr(\rho_S)=\Tr(\rho_{S}')$ and if $\rho_S\geq 0$ then $\rho_S'$ is also positive semi-definite. Thus, we see that the map, $\mathcal{E}$, preserves positivity as well as the trace. Actually, not only the map $\mathcal{E}$ preserves positivity, but also any map of the form $I_d\otimes\mathcal{E}$ does the same for all $d-$dimensional identity operators, $I_d$. Thus, $\mathcal{E}$ is a CPTP map. The map $\mathcal{E}$ is completely positive, because we started with a separable initial state, $\rho_S\otimes\rho_E$. If instead of $\rho_S\otimes\rho_E$, we would start with an initial entangled state, $\rho_{SE}$, acting on the joint Hilbert space, $\mathcal{H}_S\otimes\mathcal{H}_E$, then the corresponding map acting on the system $S$ would still be positive but not completely positive. We would like to note here that the complete positivity of the map, $\mathcal{E}$, depends on the non-existence of quantum correlation in the initial state between $S$ and $E$, and not on the classical correlation. That is, even if a classical mixture of product states, $\{\rho_S^i\otimes\rho_E^i\}_i$, had been considered in place of the product state $\rho_S\otimes\rho_E$, the map $\mathcal{E}$ would still have been completely positive. 

In the seminal paper by Frey \emph{et al}.~\cite{Frey}, energy extraction from shared batteries by operating CPTP maps on a single party has been discussed. Precisely, considering a bipartite quantum system, characterized by a state $\rho_{AB}$ and a Hamiltonian $H_{AB}$, acting on the joint Hilbert space $\mathcal{H}_A \otimes \mathcal{H}_B$, energy extraction from the system using LCPTP maps of the form $\mathcal{E}_A\otimes I_B$ has been explored, where $I_B$ is the identity operator acting on $\mathcal{H}_B$. The maximum energy extractable using such operations can be written as
\begin{eqnarray}
    \Delta W^{CP}_{\max}(H_{AB},\rho_{AB})=\text{Tr}(H_{AB} \rho_{AB})\nonumber\\ - \min_{\mathcal{E}_A} \text{Tr}\left[H_{AB} (\mathcal{E}_A \otimes \mathcal{I}_B) \rho_{AB}\right].
\end{eqnarray}
Here the minimization is performed over all CPTP maps, $\mathcal{E}_A$. 

If no energy can be extracted using CPTP processes of the type $\mathcal{E}_A\otimes I_B$, i.e., $W^{CP}_{\max}= 0$, then the state, $\rho_{AB}$, is said to be strong local (CPTP-local) passive with respect to the Hamiltonian, $H_{AB}$.
The concept of strong local passivity was first introduced in Ref.~\cite{Frey}. Later in 2019, a necessary and sufficient condition of CPTP-local passivity was derived~\cite{Alhambra}.
Explicitly, in Ref.~\cite{Alhambra}, the authors have shown that a state, $ \rho_{AB}$, corresponding to a Hamiltonian, $H_{AB}$, is CPTP-local passive, if and only if $\text{Tr}_{A'}\left (d_A^2 \ketbra{\phi^+}{\phi^+} C_{AA'} \right )$ is Hermitian and
\begin{equation}
\label{Al_CP_passive}
    C_{AA'} - \text{Tr}_{A'}\left (d_A^2 \ketbra{\phi^+}{\phi^+} C_{AA'} \right ) \otimes \mathcal{I}_{A'} \ge 0.
\end{equation}
Here $A'$ represents another system, described by the Hilbert space $\mathcal{H}_{A'}$ which is a copy of $\mathcal{H}_A$. The identity operators acting on $\mathcal{H}_A$ and $\mathcal{H}_{A'}$ can be denoted as $I_A$ and $I_{A'}$, respectively. $\kett{\phi^+}=1/d_A\sum_i \kett{i}\kett{i}$ is a maximally entangled state of $\mathcal{H}_A\otimes\mathcal{H}_{A'}$, $d_A$ is the dimension of $\mathcal{H}_A$ or $\mathcal{H}_{A'},$ and $C_{AA'}$ is $\text{Tr}_B \left [(\rho_{AB}^{\text{T}_A} \otimes {I}_{A'}) ({I}_{A} \otimes H_{BA'})\right ]$, where $\Tr_B$ and $\text{T}_A$ represents tracing out system $B$ and transposition on system $A$, respectively. $H_{BA'}$ is the same as $H_{AB}$ with the only differences being that the order of the particles in $H_{BA'}$ is just the opposite of the same in $H_{AB}$, and $H_{BA'}$ acts on $\mathcal{H}_B\otimes \mathcal{H}_{A'}$ instead of $\mathcal{H}_A\otimes \mathcal{H}_{B}$.

There is a reason why one should consider CPTP-local passivity instead of CPTP passivity, i.e., the inability to extract energy from a composite state using global CPTP operations~\cite{Alhambra}. For any given Hamiltonian, it is always possible, via global CPTP operations, to throw away the state and prepare the system in its ground state, from which no energy can be extracted. So, if every global CPTP operation on the system is allowed, the ground state of the relevant Hamiltonian trivially becomes the only state from which no energy can be extracted. Therefore, it is more interesting to look at energy extraction using LCPTP maps. However, energy extraction using local CPTP maps for a Hamiltonian, which is also local in the same partition, will again be trivial because of the same reason. So, the non-trivial situations arise where the CPTP maps act on a subsystem and a non-local Hamiltonian describes the entire system, as  considered in Ref.~\cite{Alhambra} and also here. 

\section{Comparison between CPTP-open and NCPTP-open batteries}
\label{sec:3}
\subsection{Extraction of energy from CPTP-local passive states}
\label{ncp1}
Here we discuss the extraction of energy from state, $\widetilde{\rho}_{AB}=p_1 \ketbra{\psi^+}{\psi^+} + (1-p_1) \ketbra{\psi^-}{\psi^-}$, using LNCPTP maps within the range of $p_1$ for which $\widetilde{\rho}_{AB}$ is CPTP-local passive. In this regard, we consider $\widetilde{\rho}_{AB}$ to be a part of a larger  system, with its state being given by
$\kett{\widetilde{\Psi}}_{EAB}=\sqrt{p_1}\kett{\xi}\kett{\psi^+}+\sqrt{1-p_1}\kett{\xi^\perp}\kett{\psi^-}$.
Here the dimension of the environment, $E$, is considered to be 2, and $\{\kett{\xi},\kett{\xi^\perp}\}$ is an orthonormal basis of the environment's Hilbert space.
$\widetilde{\rho}_{EAB}=\kett{\widetilde{\Psi}}\braa{\widetilde{\Psi}}_{EAB}$ is therefore, in general, entangled in the bipartition $E:AB$, and the amount of entanglement depends on $p_1$. Now if we apply a global unitary on $EA$, i.e., $U_{EA}\otimes {I}_B$, and trace out $E$ from the entire final state, the change of $\widetilde{\rho}_{AB}$ will, in general, be governed by an LNCPTP map. The maximum extractable energy in such a process is  
given in a similar manner as in Eq.~(7) of the main text,
but $\rho_{AB}$ and $H_{AB}$ replaced by $\widetilde{\rho}_{AB}$ and $\widetilde{H}_{AB}$ respectively. Let us denote this quantity by $\Delta W_{p}^{NCP} (\widetilde{\rho}_{AB},\widetilde{H}_{AB})$.

An arbitrary two-qubit global unitary  can be expressed in the form following form~\cite{Kraus}:
\begin{eqnarray}
\label{UEA}
      U_{EA} = U_E \otimes U_A U_d V_E \otimes V_A, 
\end{eqnarray}
where \(U_E, V_E, U_A, V_A \in U(2)\).
$U_E$, $U_A$, $V_E$, $V_A$ are  local
unitaries and \(U_d\) is, in general, a ``non-local'' unitary on \(\mathbb{C}^2 \otimes \mathbb{C}^2\) given by
\begin{equation}
\label{Ud}
U_d = \exp(-i\eta_x \sigma^x \otimes \sigma^x -i\eta_y \sigma^y \otimes \sigma^y -i\eta_z \sigma^z \otimes \sigma^z). 
\end{equation}
Here \(\eta_x\), \(\eta_y\), \(\eta_z\) are real parameters lying in the range \(0 \le \eta_x,~ \eta_y,~ \eta_z \le \pi/2 \).
In Eq.~\eqref{UEA}, the local unitaries are of the form
\begin{equation}
\label{unitary1}
        \mathbb{U}_{k}=
        \left( {\begin{array}{cc}
        \cos\frac{ \alpha_k}{2}e^{\frac{i}{2}( \beta_k+\gamma_k)} &  \sin\frac{ \alpha_k}{2}e^{-\frac{i}{2}( \beta_k-\gamma_k)} \\
        -\sin\frac{ \alpha_k}{2}e^{\frac{i}{2}( \beta_k-\gamma_k)} &  \cos\frac{ \alpha_k}{2}e^{-\frac{i}{2}( \beta_k+\gamma_k)} 
        \end{array} } \right),
\end{equation}
where $k=1$, 2, 3, and 4 represent, respectively, $U_A$, $U_E$, $V_A$, and $V_E$. Here $ \alpha_k \in [0,\pi]$, $ \beta_k \in [0,4\pi)$ and  $\gamma_k \in [0,2\pi]$. 

Let us consider the set of  $\widetilde{\rho}_{AB}$ which obeys $p_1 \le 1/2$ and $r \ge p/(1-2p_1)$. Since these states are CPTP-local passive, $\Delta W^{CP}_{\max}=0$. Considering many such $\widetilde{\rho}_{AB}$ we determine $\Delta W^{NCP}_{p}$ 
The maximization over the unitaries $U_{EA}$ is performed by expressing $U_{EA}$ in terms of the parameters $\alpha_k$, $\beta_k$, $\beta_k$, $\eta_x$, $\eta_y$, $\eta_z$ and maximizing the extractable energy over these parameters. The optimization has been numerically performed using the algorithm, ISRES, available in NLOPT, which is a nonlinear optimization package. The numerical accuracy of the results is maintained up to the second decimal place. Since the maximization takes care of all local and global unitaries, $U_{EA}$, $\Delta W^{NCP}_{p}$ does not depend on
the Schmidt basis,
$\{\kett{\xi},\kett{\xi^\perp}\}$, of the environment, $E$. 

The behavior of 
$\Delta W^{NCP}_{p}(\widetilde{\rho}_{AB},\widetilde{H}_{AB})$, is depicted in Fig.~1 of the main text with respect to the entanglement, $S$, of $\widetilde{\rho}_{EAB}$ in the bipartition $E:AB$, for different values of interaction strength, $r$.  We measure $S$ using entanglement entropy. We obtain that $\Delta W^{NCP}_{p}$ is zero at $S=0$. This is evident as $\widetilde{\rho}_{EAB}$ is a pure product state when $S$ is zero, and the generated map is still completely positive. 
Further, we realize that with increase in entanglement of the initial system-environment state, extractable energy using the LNCPTP evolution also increases. Hence, we can conclude that energy extraction using LNCPTP dynamics is potentially useful and advantageous over energy extraction using LCPTP processes. Therefore, a state that seemed to be ``passive" under LCPTP maps is no longer ``passive" if a more general positive map, the physical LNCPTP maps, are allowed. Note that by ``passive," we mean that the state is unable to provide any energy through the application of the corresponding map under consideration. Moreover, we observe that the maximum energy extracted using the LNCPTP operation, $\Delta W_p^{NCP}$, increases with increasing strength of the interaction, $r$. 

The following important features
\begin{itemize}
\item $\Delta W^{CP}_{\max}=0$ for all CPTP-local passive states and 
\item $\Delta W^{NCP}_{\max}\geq \Delta W^{NCP}_{p}>0$ for a large set of CPTP-local passive states
\end{itemize}
can be utilized to formulate a witness to the non-complete positivity of the governing dynamics.
That is, to check if an unknown map is completely positive or not, we can take a CPTP-local passive state and try to extract energy from it by operating the unknown map on a particular subsystem of the state on which, if CPTP operations are applied, no energy can be extracted. If a non-zero amount of energy can be extracted by operating the map, then the map is certainly non-CPTP. However, if the energy extraction is still zero, then nothing can be inferred about the non-CPTP behavior of that unknown map. Non-zero energy extraction from a CPTP-local passive state not only detects NCPTP maps but also entanglement shared between $E$ and $AB$. It is to be borne in mind, however, that in the process of generation of the NCPTP dynamics, we here restrict only to pure $\widetilde{\rho}_{EAB}$. The maximum extractable energy can in principle be higher if a more general $\widetilde{\rho}_{EAB}$ is considered, which might be mixed as well, keeping $\widetilde{\rho}_{AB}$ fixed.

We have shown a CPTP-local passive state remains CPTP-local passive even if an asymptotically large number of copies of the state are considered. Therefore we can conclude that the gap in the energy extraction using LCPTP and LNCPTP maps will not decrease even if we move towards the asymptotic limit, at least in the case of CPTP-local passive states.

\begin{figure*}
\centering
\includegraphics[height=3
cm]{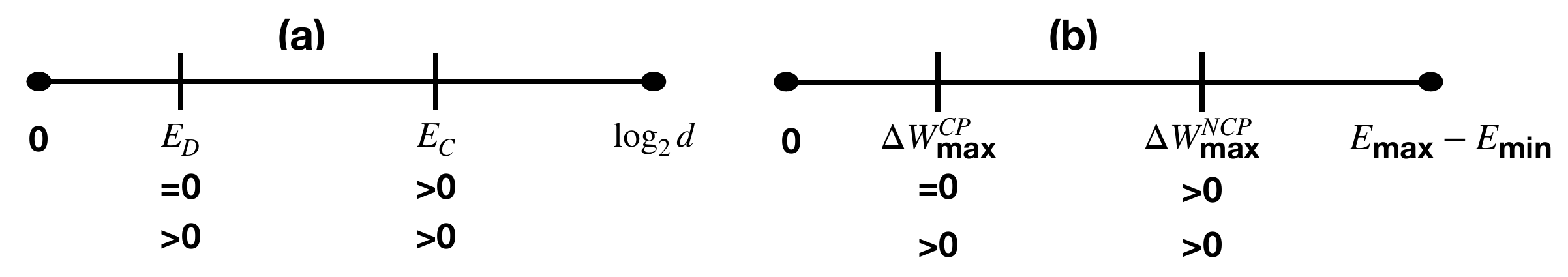}%
\caption{Analogy in the ordering of distillable entanglement and entanglement cost with that of the maximum energy extracted from CPTP-open and NCPTP-open systems, respectively.
Panel (a) describes the fact that distillable entanglement, $E_D$, is less than or equal to the entanglement cost, $E_C$, for all quantum states in any bipartite dimensions. Thus, $E_D$ always situates itself on the left side of $E_C$ in the axis of entanglement of an arbitrary state on $\mathbb{C}^d \otimes \mathbb{C}^d$, where the left and right edges of the axis represent values 0 and log$_2d$ ebits, which are respectively the lowest and highest values of $E_D$ and $E_C$ that a state can have. Through the schematic diagram, we want to emphasize that there can be situations where $E_D=0$ but $E_C>0$ as well as situations where $E_C>E_D>0$. 
The same ordering can be found between maximum extractable energy using LCPTP processes, $\Delta {W}_{\max}^{CP}$, and LNCPTP processes, $\Delta {W}_{\max}^{NCP}$, i.e., $\Delta {W}_{\max}^{CP} \leq \Delta {W}_{\max}^{NCP}$ for all states and Hamiltonians. This is shown in 
panel (b). In case of energy extraction as well, there are examples for which $\Delta {W}_{\max}^{NCP}>\Delta {W}_{\max}^{CP}=0$ and also exemplar states for which $\Delta {W}_{\max}^{NCP}>\Delta {W}_{\max}^{CP}>0$.
The two edges of the axis shown in panel (b) represent values 0 and $E_{\max}-E_{\min}$ where $E_{\max}$ and $E_{\min}$ are respectively the highest and lowest eigenvalues of the considered Hamiltonian. Therefore, $E_{\max}-E_{\min}$ represents the maximum amount of energy that can be extracted from a state. The axes of the left and right panels have the units of ebits and energy, respectively.}
\label{entanglement}
\end{figure*}

\subsection{Finding the maximum extractable work via CPTP operations, and surpassing it using physical NCPTP maps}

In this subsection, we provide  the maximum extractable work using LCPTP operations and a certain class of states and Hamiltonians, for which extractable energy using physical NCPTP operations can cross this maximum. The maximum extractable energy using any fixed local CPTP map, $\Lambda_A\otimes I_B$, where $I_B$ is the identity operator which acts on $\mathcal{H}_B$, is given by~\cite{Alhambra}
\begin{equation}
\label{cpbound}
    \Delta W^{max}=\text{Tr}(H_{AB}\rho_{AB})-\min_{E_{AA'}} \text{Tr}(C_{AA'}E_{AA'}),
\end{equation}
where $E_{AA'}$ is the Choi matrix of the map, $\Lambda_A$, and $C_{AA'}$ is defined as previously. For CPTP maps, the eigenvalues of $E_{AA'}$, say $\lambda_i$, are all positive, and they sum up to unity, i.e., $\sum_i \lambda_i = 1$. 
Therefore, the second term of Eq.~\eqref{cpbound} involves a minimization over all hermitian operators $E_{AA'}$, under the conditions $\lambda_i>0$ and $\sum_i \lambda_i = 1$. For fixed set of eigenvalues, $\{\lambda_i\}_i$, the minimum of $\text{Tr}(C_{AA'}E_{AA'})$ will be obtained if $[C_{AA'},E_{AA'}]=0$, and the eigenvalues of $C_{AA'}$, say $c_i$, and $E_{AA'}$ obey opposite order, i.e., if $c_i<c_j$ then $\lambda_i\ge\lambda_j$~\cite{Lenard}. Thus we have $$\min_{E_{AA'}} \text{Tr}(C_{AA'}E_{AA'})=\min_{\{\lambda_i\}}\sum_i c_i\lambda_i.$$
We have to now optimize $\sum_i c_i\lambda_i$ over $\lambda_i$ under the constraints that, if $c_i>c_j$, then $\lambda_i\le\lambda_j$, and $\sum_i \lambda_i = 1$. 
Let us now restrict ourselves to two-qubit batteries. In the case of two-qubit batteries, the function needed to be optimized would consist of four terms, i.e., 
\begin{eqnarray}
    F &=& \sum_{i=1}^4c_i \lambda_i \nonumber \\
      &=& c_4 +\lambda_1(c_1-c_4)+\lambda_2(c_2-c_4)+\lambda_3(c_3-c_4).\nonumber \;\;\;
\end{eqnarray}
Here we have used the relation, $\lambda_4=1-\lambda_1-\lambda_2-\lambda_3$. Since the function, $F$, as well as the constraints on $\{\lambda_i\}$ are linear, $F$ will attain its optimum value at the edges of the feasible region. Without loss of generality, let $c_1<c_2<c_3<c_4$. Then the edges of the feasible region of $(\lambda_1,\lambda_2,\lambda_3)$ where all constraints (i.e., $\lambda_i\geq \lambda_{i+1}\geq 0$ for all $i$) will be satisfied, correspond to $(1,0,0)$ and $(0,0,0)$, which reduces to $F=c_1$ and $c_4$ respectively. Since the smallest among them is $c_1$, therefore  $F_{min}=c_1$. Proceeding in this manner, it can be shown for arbitrary system dimensions that the second quantity on the right-hand side of Eq.~\eqref{cpbound} is always equal to $c_{min}$, where $c_{min}$ is the minimum eigenvalue of $C_{AA'}$. Therefore the maximum amount of work that can be extracted from a system using LCPTP maps is 
\begin{equation}
\label{cpmax}
\Delta W^{CP}_{max}=\text{Tr}(H_{AB}\rho_{AB})-c_{min}.
\end{equation}
In this derivation, we have assumed that all eigenvalues of $C_{AA'}$ are non-degenerate. Thus if $C_{AA'}$ has degenerate eigenvalues, Eq.~\eqref{cpmax} may not be valid.

Let us now find the class of physical NCPTP maps which outperform $\Delta W^{CP}_{max}$. We consider work extraction using physical NCPTP maps, where we restrict $U_{EA}$ to controlled unitaries of the form $U_{EA}=U_C=\ketbra{0}{0}\otimes \mathbb{I}_2+\ketbra{1}{1}\otimes \mathbb{U}_2$. Here $\mathbb{I}_2$ is the single-qubit identity operator, and $\mathbb{U}_2 \in SU(2)$ which has the same form as given in Eq.~\eqref{unitary1}. We choose the set of two-qubit states $\widetilde{\rho}_{AB}$, which is pure in an extended Hilbert space. We consider the same class of non-local Hamiltonian, $\widetilde{H}_{AB}$. To reduce the number of free parameters, we fix $p_1=1/4$. It is an arbitrary choice, and the following calculations can be easily altered accordingly for other values of $p_1$.
From Eq.~\eqref{cpmax}, we get the maximum extractable work using CPTP maps for this class of $\widetilde{\rho}_{AB}$ and $\widetilde{H}_{AB}$, that is 
\begin{equation}
\label{wcpmax}
    W^{CP}_{max}=\frac{1}{2}(p+q-2r).
\end{equation}
Work extracted by the class of physical NCPTP maps generated using the controlled unitary $(U_C)$ is given by
\begin{equation}
\label{wncp_uc}
    W^{NCP}_{U_C}=\frac{3}{8}rx,
\end{equation}
where $x=\cos(\alpha_2-\beta_2-\gamma_2)+2\cos(\beta_2+\gamma_2)+\cos(\alpha_2+\beta_2+\gamma_2)-4$. Here $\alpha_2$, $\beta_2$ and $\gamma_2$ are the parameters of the unitary, $ \mathbb{U}_2$ [see Eq.~\eqref{unitary1}]. Hence $-8\le x \le 0$. 
Comparing equations ~\eqref{wcpmax} and ~\eqref{wncp_uc}, we find that the maximum extractable work using LCPTP maps will be surpassed by LNCPTP operations, for a class of states, $\widetilde{\rho}_{AB}$, and Hamiltonians, $\widetilde{H}_{AB}$, if $3rx/8 > (p+q-2r)/2$.
Let us look at the condition more closely.
If we restrict to $r<0$ and $x\ne 0$, the condition reduces to
\begin{equation}
\label{condition}
    \frac{3}{8} |r||x| > |r|+(p+q)/2.
\end{equation}
If $p+q < 0$, then the right-hand side of the inequality~\eqref{condition} becomes smaller than $ |r|$. Moreover, if $|x| > 8/3$, then the left-hand side of inequality~\eqref{condition} becomes greater than $ |r|$. Therefore under such conditions, inequality~\eqref{condition} gets satisfied, implying that by using these classes of NCPTP operations and the relevant states and Hamiltonians, one can extract more energy than using local-CPTP operations.
This is an explicit situation of energy extraction from CPTP-non passive states.
Next, if we consider $r=(p+q)/2$, then the set of states turn out to be local-CPTP passive corresponding to the Hamiltonian, $\widetilde{H}_{AB}$. In this case, $p+q<0$ will provide the condition for extracting energy using NCPTP operations, hence depicting an instance of the possibility of energy extraction from CPTP-local passive states.

\subsection{Energy extraction from CPTP non-passive battery states using  LNCPTP maps}
\label{cp_ncp}
The entanglement of an arbitrary $d$-dimensional state, if measured using distillable entanglement or entanglement cost, can vary from zero to $\log_2d$. Along the axis of entanglement, as shown in Fig.2-(a) of SM, distillable entanglement, $E_D$, always lies to the left of entanglement cost, $E_C$, that is, $E_D$ is always less or equal to $E_C$. There exist examples for which $E_D=0$ but $E_C>0$~\cite{Vidal_Cirac} which refers to the bound entangled states. There also exist situations for which $0<E_D<E_C$~\cite{Vidal_Cirac_pra}. This implies that the state contains a non-zero part of the entire entanglement that can be distilled, but there is still some entanglement bound within the state that cannot be extracted. We can draw an analogy between these properties and the extractable energies from a battery using LCPTP or LNCPTP operations. We determined that there exist states, $\widetilde{\rho}_{AB}$, and Hamiltonians, $\widetilde{H}_{AB}$, for which $\Delta W_{\max}^{CP}(\widetilde{\rho}_{AB},\widetilde{H}_{AB})=0$ but $\Delta W_{\max}^{NCP}(\widetilde{\rho}_{AB},\widetilde{H}_{AB})\geq \Delta  W_p^{NCP}(\widetilde{\rho}_{AB},\widetilde{H}_{AB})>0$, as depicted in Fig.2-(b) of SM. This implies that these CPTP-local passive states consist of some ``bound" energy that gets released when LNCPTP maps are applied. Motivated by this feature, we try to see if the energy extracted by LNCPTP processes and that from an optimum CPTP process can bear a property similar to $E_D>E_C>0$. Specifically, we want to examine if a non-CPTP-passive state can still have some energy bound within the state, which can only be freed when one considers physical LNCPTP maps in the set of allowed operations.

Let us fix the Hamiltonian to $\widetilde{H}_{AB}=p(\sigma^z \otimes \mathcal{I}) + q(\mathcal{I} \otimes \sigma^z) + r(\sigma^x \otimes \sigma^x + \sigma^y \otimes \sigma^y)$  and maximize the extractable energy using LNCPTP maps, $\Delta W_{p}^{NCP}(\Bar{\rho}_{AB},\widetilde{H}_{AB})$ over all rank-two battery states, $\Bar{\rho}_{AB}$. To construct the states, first we consider the following pure state
\begin{eqnarray}
\kett{\psi}&=&a\kett{00}+b\kett{01}+c\kett{10}+d\kett{11}+e\kett{20}+f\kett{21} \nonumber \\
 &+&g\kett{30}+h\kett{31},
\end{eqnarray}
and then trace out the second party. The fact that the dimension of the second party is considered to be two, ensures that the resulting state, $\Bar{\rho}_{AB}=\Tr_2(\kett{\psi}\braa{\psi})$, is rank-two. Here the parameters $a$, $b$, $c$, $d$, $e$, $f$, $g$, and $h$ are arbitrary complex numbers chosen in such a way that $\kett{\psi}$ is normalized. 

A discussion on how we construct the LNCPTP maps is provided here.
Let us first fix the Hamiltonian to $\widetilde{H}_{AB}$  and maximize the extractable energy using LNCPTP maps, $\Delta W_{p}^{NCP}(\Bar{\rho}_{AB},\widetilde{H}_{AB})$ over all rank-two battery states, $\Bar{\rho}_{AB}$. We consider $\Bar{\rho}_{AB}$ to be a part of a bigger space $EAB$ which consists of the battery $AB$ as well as an environment $E$. The complete environment-battery state is pure and this pure state can be expressed as $\kett{\Bar{\Psi}}_{EAB}=\sqrt{x_1}\kett{0\nu_1}+\sqrt{x_2}\kett{1\nu_2}$, where $x_1$ and $x_2$ are the two largest eigenvalues of $\Bar{\rho}_{AB}$, and their corresponding eigenvectors are $\kett{\nu_1}$ and $\kett{\nu_2}$. $\{\kett{0},\kett{1}\}$ is the set of Schmidt vectors of the environment. If both $x_1$ and $x_2$ are non-zero, then $\kett{\Bar{\Psi}}_{EAB}$ would be entangled in the bipartition $E:AB$. Thus a global unitary operation on the entire state of $\kett{\Bar{\Psi}}_{EAB}$ can reduce to an NCPTP operation on the battery $\Bar{\rho}_{AB}$ which would make ${AB}$ is an NCPTP-open quantum battery. Instead of global NCPTP-operations, in this work, we only consider LNCPTP maps. To generate such maps, we act unitaries of the form $U_{EA}\otimes {I}_B$ on $\kett{\Bar{\Psi}}_{EAB}$ and trace out the environment, $E$. This operation allows for the most general LNCPTP map on $\Bar{\rho}_{AB}$. The corresponding maximum energy extractable through this evolution is given by $\Delta \Bar{W}_{M}^{NCP}(r) =\max_{\Bar{\rho}_{AB}}\Delta {W}_p^{NCP}(\Bar{\rho}_{AB},\widetilde{H}_{AB})$ where $\Bar{\rho}_{EAB}=\kett{\Bar{\Psi}}\braa{\Bar{\Psi}}_{EAB}$. Here we have maximized not only over the LNCPTP channels but also on the initial states, $\Bar{\rho}_{EAB}$. For various values of the Hamiltonian parameter, $r$, we numerically obtained the optimal extractable energy using LNCPTP maps, i.e., $\Delta \Bar{W}_{M}^{NCP}(r)$, and the corresponding state, $\Bar{\rho}_{AB}^r$, which acquires this optimal value.  

From the discussions of the main paper, we find that on a similar footing with Fig.~\ref{entanglement}-(a), we have found that when $\Delta {W}_{\max}^{CP}=0$, i.e., for CPTP-local passive states, $\Delta {W}_{p}^{NCP}$ can be positive. Moreover, even when $\Delta {W}_{\max}^{CP}>0$, we have seen that $\Delta {W}_{p}^{NCP}$ can also be positive and much greater than $\Delta {W}_{\max}^{CP}$. 
This fact is depicted in Fig.~\ref{entanglement}-(b).
We have considered only those LNCPTP maps that can be generated by applying unitaries on the pure state $\kett{\Bar{\Psi}}_{EAB}$. The situation can be generalized by considering mixed states of the entire system, $EAB$, which may further enhance the extractable energy. It can be inferred that $\Delta {W}_{\max}^{CP}$ and $\Delta {W}_{\max}^{NCP}$ exhibit relation similar to that between distillable entanglement, $E_D$, and entanglement cost, $E_C$, respectively.

It should be noted that if instead of rank-two states, rank-one states were considered in place of $\Bar{\rho}_{AB}$ or $\widetilde{\rho}_{AB}$, the energy extraction operation would reduce to a CPTP map. To get the flavor of NCPTP maps, one must assure that the rank of the battery state is higher than unit. For numerical simplicity, we have restricted ourselves to rank-two states. We expect that considering higher rank states would not affect the results qualitatively.

\vspace{0.2cm}

\section{Necessary and sufficient condition for NCPTP-local passivity}
\begin{theorem}
\label{NCPTP-local-passivity}
The pair, $\{H_{AB}, \rho_{EAB}\}$ is NCPTP-local passive with respect to subsystem $A$ if and only if 
$C-C'\ge 0$,
where $ \braket{\alpha\beta|C'|\alpha' \beta'}=\delta_{\alpha \alpha'} \sum_i \braket{ii|C|\beta' \beta}$,
$\forall$ $\alpha, \beta, \alpha'$ and $\beta'$. Here  $\ket{i} \in \mathcal{H}_{E_i} \otimes \mathcal{H}_{A_i}$, $\ket{\beta^{(')}} \in \mathcal{H}_{E_{1(2)}} \otimes \mathcal{H}_{A_{1(2)}}$, and $C\equiv C_{E_1 A_1 E_2 A_2}=\bra{\phi}\left(\mathcal{I}_{E_1}\otimes H_{A_1 B_1}\otimes\rho_{E_2 A_2 B_2}^T \right) \ket{\phi}_{B_1 B_2}$, where $\ket{\phi}_{B_1B_2}$ is the unnormalized maximally entangled state, given by $\ket{\phi}_{B_1B_2}=\sum_j \ket{jj}$.
\end{theorem}

\noindent \textit{Proof.}
The maximum amount of energy that can be extracted from a battery state, $\rho_{AB}$, using physically realizable local NCPTP maps is
\begin{eqnarray}
\label{ncp_max}
  \Delta W_{\max}^{NCP}(\rho_{AB},H_{AB})=  \text{Tr}\left[\rho_{AB} H_{AB} \right] \nonumber\\-\min_{\Lambda^N_A} \text{Tr}\big[\Lambda^N_A\otimes I_B(\rho_{AB}) H_{AB} \big]. 
\end{eqnarray}
In Eq.\eqref{ncp_max}, the minimization is performed over all NCPTP-maps, $\Lambda^N_A$. The subscript and superscript, $A$ and $N$, used in the notation of the map, denote, respectively, that the map acts on the subsystem $A$, and the map may not be completely positive.
Here, we refer to those NCPTP operations that are valid quantum maps, i.e., which are physically realizable. 
To construct such physically realizable NCPTP maps, consider a tripartite system consisting of a bipartite battery, $AB$, and an environment $E$. If the tripartite initial state, $\rho_{EAB}$, is entangled within the bipartition $E:AB$, then the battery state, $\rho_{AB}$, which can be obtained by evolving $\rho_{EAB}$ by a global unitary, and subsequently tracing out the environment $E$, 
will obey NCPTP evolution. We want to evaluate the maximum extractable energy from $\rho_{AB}$ for a given Hamiltonian $H_{AB}$, using such NCPTP operations on the system $A$. 
Therefore, the maximum extractable energy from $\rho_{AB}$, with respect to Hamiltonian $H_{AB}$, using such kind of LNCPTP operation is given by 
\begin{eqnarray}
\label{ncp_work}
   && \Delta W^{NCP} (\rho_{AB},H_{AB})=  \text{Tr}\left[\rho_{AB}  H_{AB} \right] \nonumber \\ 
   &-& \min_{U_{EA}} \text{Tr}\big[\text{Tr}_E(U_{EA}\otimes {I}_B\rho_{EAB}U_{EA}^{\dagger}\otimes {I}_B) H_{AB} \big], \;\;\;\;\;\;
\end{eqnarray}
where $U_{EA}$ are global unitaries acting on $E$ and $A$.
Now the second term on the right hand side of Eq.~\eqref{ncp_work} can be simplified as follows
\begin{eqnarray}
\label{lambda}
   F &=& \text{Tr}\big[\text{Tr}_E(U_{EA}\otimes {I}_B\rho_{EAB}U_{EA}^{\dagger}\otimes {I}_B) H_{AB} \big] \nonumber \\
    &=& \Tr\left[\left(U_{EA}\otimes {I}_B\rho_{EAB}U_{EA}^{\dagger}\otimes {I}_B\right)\left(\mathcal{I}_E \otimes H_{AB} \right)\right] \nonumber \\
    &=& \Tr\left[\Lambda(\rho_{EAB})\left(\mathcal{I}_E \otimes H_{AB} \right)\right] ,
\end{eqnarray}
where $\Lambda(\chi_{EAB}) = U_{EA}\otimes {I}_B\chi_{EAB}U_{EA}^{\dagger}\otimes {I}_B$. The equation~\eqref{lambda} can be further rewritten as 
\begin{eqnarray}
\label{lambda1}
   F &=& \Tr\left[\Lambda(\rho_{EAB})\left(\mathcal{I}_E \otimes H_{AB} \right)\right] \nonumber \\
    &=& \Tr\left(\widetilde{C}_{E_1A_1B_1E_2A_2B_2}\widetilde{E}_{E_1A_1B_1E_2A_2B_2}\right),
\end{eqnarray}
where $\widetilde{C}_{E_1A_1B_1E_2A_2B_2}=\mathcal{I}_{E_1}\otimes H_{A_1B_1}\otimes\rho_{E_2A_2B_2}^T$ and $\widetilde{E}_{E_1A_1B_1E_2A_2B_2}=\sum_{i,j,k,l,m,n}\Lambda \left(\ketbra{ijk}{lmn}\right)\otimes\ketbra{ijk}{lmn}$. Here, $H_{A_1B_1}$ and $\rho_{E_2A_2B_2}$ represent the Hamiltonian and state 
belonging to the Hilbert spaces $\mathcal{H}_{A_1}\otimes\mathcal{H}_{B_1}$ and $\mathcal{H}_{E_2}\otimes\mathcal{H}_{A_2}\otimes\mathcal{H}_{B_2}$ respectively, where $\mathcal{H}_{E_i}$, $\mathcal{H}_{A_i}$ and $\mathcal{H}_{B_i}$ denote the Hilbert spaces of the environment, and subsystems $A$ and $B$ respectively for $i=1$ to $2$. $\widetilde{E}_{E_1A_1B_1E_2A_2B_2}$ is essentially the Choi state corresponding to the action of 
local unitaries $U_{EA}$. So Eq.~\eqref{lambda1} can be finally simplified to
\begin{eqnarray}
F &=& \Tr\left(\widetilde{C}_{E_1A_1B_1E_2A_2B_2}\widetilde{E}_{E_1A_1B_1E_2A_2B_2}\right) \nonumber \\
&=& \Tr\left[C_{E_1A_1E_2A_2} \ketbra{\Psi}{\Psi}\right], 
\end{eqnarray}
where $C_{E_1 A_1 E_2 A_2}=\bra{\phi}\left(\mathcal{I}_{E_1}\otimes H_{A_1 B_1}\otimes\rho_{E_2 A_2 B_2}^T \right) \ket{\phi}_{B_1 B_2}$, and $\ketbra{\Psi}{\Psi}$ is the Choi state corresponding to the unitary operator, $U_{E_1 A_1}$, and therefore is pure.
The explicit form of $\ket{\Psi}=U_{E_1 A_1}\otimes I_{E_2 A_2} \sum_{i,j}\ket{ij}\otimes\ket{ij}$.
Moreover, since our proof is independent of the dimension of the local subsystems, for simplicity, we denote $\widetilde{A}\equiv A_1 B_1$ and $\widetilde{A}'\equiv A_2 B_2$. Therefore $F=\sum_{i,j}\Tr(C_{\widetilde{A}\widetilde{A}'} U_{\widetilde{A}} \otimes I_{\widetilde{A}'}\ket{\tilde{j}\tilde{j}}\bra{\tilde{i}\tilde{i}} U^{\dagger}_{\widetilde{A}} \otimes I_{\widetilde{A}'})$, where $\ket{\tilde{j}}\equiv\ket{ij}$ and $\ket{\tilde{i}}\equiv\ket{lm}$. We will omit the subscripts further
in our analysis.

Our goal is to minimize the quantity, $F=\sum_{i,j}\Tr\left[\left(U \ketbra{\tilde{j}}{\tilde{i}}U^{\dagger}\otimes \ketbra{\tilde{j}}{\tilde{i}}\right) C\right]$, with respect to the unitaries, $U$.
Let us look at $F$ more closely.
\begin{eqnarray}
    F &=& \sum_{i,j} \Tr\left(U \ketbra{\tilde{j}}{\tilde{i}}U^{\dagger} \braket{\tilde{i}|C|\tilde{j}} \right) \nonumber \\
    &=& \sum_{i,j} U_{xp} \ketbra{\tilde{j}_p}{\tilde{i}_q}U^{\dagger}_{qy} \braket{\tilde{i}|C|\tilde{j}}_{yx}
\end{eqnarray}
Since we optimize over the bases $\ket{\tilde{j}}$ by using global unitaries, it is enough to consider $\ket{\tilde{j_p}}=\delta_{jp}$. This gives
\begin{eqnarray}
    F &=& \sum_{ij}U_{xj}U^*_{yi}\braket{\tilde{i}|C|\tilde{j}}_{yx} \nonumber \\
   &=& \sum_{ij}(R_{xj}+iM_{xj})(R_{yi}-iM_{yi})\braket{\tilde{i}|C|\tilde{j}}_{yx},
\end{eqnarray}
where $R$ and $M$ denotes the real and imaginary parts of $U$. Therefore, we have to minimize $F$ over $R$ and $M$ subject to the constraint $\sum_j(R_{ji}-iM_{ji})(R_{jk}+iM_{jk})=\delta_{ik}$. The Lagrangian corresponding to the constrained minimization problem is given by
\begin{eqnarray}
    L &=& \sum_{ij}(R_{xj}+iM_{xj})(R_{yi}-iM_{yi})\braket{\tilde{i}|C|\tilde{j}}_{yx} \nonumber \\
    &-& \sum_{ik}\lambda_{ik}\left[\sum_j(R_{ji}-iM_{ji})(R_{jk}+iM_{jk})-\delta_{ik}\right], \;\;\;\;\;\;
\end{eqnarray}
where $\lambda_{ik}$ are the respective Lagrange multipliers. The first order necessary condition of obtaining minimum of the Lagrangian, $L$, is $\partial L/\partial R_{\alpha \beta}=0$ and $\partial L/\partial M_{\alpha \beta}=0$, $\forall$ $\alpha, \beta$~\cite{Magnus}. These two conditions are explicitly given by
\begin{eqnarray}
    \sum_{iy}(R_{yi}-iM_{yi})\braket{\tilde{i}|C|\beta}_{y\alpha} &-& \sum_k\lambda_{\beta k}(R_{\alpha k}+iM_{\alpha k}) + \nonumber \\
     \sum_{xj}(R_{xj}+iM_{xj})\braket{\beta|C|\tilde{j}}_{\alpha x} &-& \sum_i\lambda_{i \beta}(R_{\alpha i}-iM_{\alpha i}) =0 \nonumber
\end{eqnarray}
and
\begin{eqnarray}
    \sum_{iy}(R_{yi}-iM_{yi})\braket{\tilde{i}|C|\beta}_{y\alpha} &+& \sum_k\lambda_{\beta k}(R_{\alpha k}+iM_{\alpha k}) - \nonumber \\
     \sum_{xj}(R_{xj}+iM_{xj})\braket{\beta|C|\tilde{j}}_{\alpha x} &-& \sum_i\lambda_{i \beta}(R_{\alpha i}-iM_{\alpha i}) =0 \nonumber 
\end{eqnarray}
respectively. Our aim is to find a necessary condition of NCPTP-local passivity. This statement is equivalent to finding the conditions under which the unitary, $U_{\widetilde{A}}=I_{\widetilde{A}}$, gives the minimum.
Therefore, using the first order necessary conditions of minimum, the following two equations are obtained
\begin{eqnarray}
    \lambda_{\alpha \beta} = \sum_i \braket{\tilde{i}\tilde{i}|C|\alpha \beta}, \;\;\;\;
    \lambda_{\beta \alpha} = \sum_i \braket{\alpha \beta|C|\tilde{i}\tilde{i}}.
\end{eqnarray}
Now, the second order necessary condition for attaining the minimum of the Lagrangian, $L$, is that, in addition to satisfying the first order conditions, its Hessian is positive semidefinite~\cite{Magnus}, i.e. $\mathbb{H}\ge0$, where the Hessian of $L$, $\mathbb{H}$, is given by
\begin{eqnarray}
    \mathbb{H}=\left( {\begin{array}{cc}
        \frac{\partial^2L}{\partial R_{\alpha'\beta'}\partial R_{\alpha \beta}} & \frac{\partial^2L}{\partial R_{\alpha'\beta'}\partial M_{\alpha \beta}} \\
        \frac{\partial^2L}{\partial M_{\alpha'\beta'}\partial R_{\alpha \beta}} & \frac{\partial^2L}{\partial M_{\alpha'\beta'}\partial M_{\alpha \beta}}
        \end{array} } \right).
\end{eqnarray}
The second order derivatives of $L$ which also satisfy the first order necessary conditions are given as follows. The diagonal blocks of $\mathbb{H}$ are equal and are given by
\begin{eqnarray}
\label{one}
 \braket{\alpha\beta|X|\alpha'\beta'} &=&  \frac{\partial^2L}{\partial R_{\alpha'\beta'}\partial R_{\alpha \beta}} = \frac{\partial^2L}{\partial M_{\alpha'\beta'}\partial M_{\alpha \beta}} \nonumber \\
  &=& \braket{\alpha'\beta'|C|\alpha \beta}+\braket{\alpha\beta|C|\alpha' \beta'} \nonumber \\
  &-& \delta_{\alpha \alpha'} \sum_i \left(\braket{\beta' \beta|C|\tilde{i}\tilde{i}}+\braket{\tilde{i}\tilde{i}|C|\beta' \beta}\right), \nonumber \\
  &=& \braket{\alpha\beta|C+C^*|\alpha' \beta'}  \nonumber \\
  &-& \delta_{\alpha \alpha'} \sum_i \braket{\tilde{i}\tilde{i}|C+C^*|\beta' \beta}, \;\;\;\;\;\;\;
\end{eqnarray}
$\forall$ $\alpha, \beta, \alpha'$ and $\beta'$. The last equation holds due to the hermiticity of the operator, $C$.
The off-diagonal blocks of the Hessian are $\partial^2L/\partial R_{\alpha'\beta'}\partial M_{\alpha \beta}=\partial^2L/\partial M_{\alpha\beta}\partial R_{\alpha' \beta'}$, which is given by
\begin{eqnarray}
\label{two}
\braket{\alpha\beta|Y|\alpha'\beta'}&=& i\braket{\alpha\beta|-C+C^*|\alpha' \beta'}  \nonumber \\
  &+& i\delta_{\alpha \alpha'} \sum_i \braket{\tilde{i}\tilde{i}|-C+C^*|\beta' \beta},\nonumber\\ &=& \braket{\alpha'\beta'|Y|\alpha\beta}\;\;\;\;\;\;\;
\end{eqnarray}
$\forall$ $\alpha, \beta, \alpha'$ and $\beta'$.
Therefore, the Hessian matrix, $\mathbb{H}$, is of 
the form
\begin{equation}
    \mathbb{H}=\left( {\begin{array}{cc}
        X & Y \\
        Y & X
        \end{array} } \right),
        \label{hessian0}
\end{equation}
where $X$ and $Y$ are as previously defined.
The conditions under which, the minimum of the Lagrangian $L$ is attained, is that, $\mathbb{H}$ is convex, which suggests that $\mathbb{H}$ is also positive semidefinite, 
and we denote this semidefiniteness by $\mathbb{H}\ge 0$. 
If a matrix is positive semidefinite, then by definition, it is hermitian and all its eigenvalues are non-negative. Therefore, $\mathbb{H}$ is also hermitian, which implies that each off-diagonal block is a null matrix, since each off-diagonal block comes out to be purely imaginary, as given in Eq.~\eqref{two}. 
Further, using generalized Schur complement~\cite{schur}, one can argue that $\mathbb{H}\ge 0$, if and only if $X\ge 0$, since we obtained $Y=0$.
Subtracting Eq.~\eqref{two} from Eq.~\eqref{one}, we get  $\braket{\alpha\beta|X|\alpha'\beta'}=\braket{\alpha\beta|C|\alpha'\beta'}-\delta_{\alpha \alpha'} \sum_i \braket{\tilde{i}\tilde{i}|C|\beta' \beta}$.
Therefore the necessary conditions for an arbitrary given pair of state and Hamiltonian to be NCPTP-local passive is
\begin{equation}
\label{necessary}
C-C'\ge 0,
\end{equation}
where $ \braket{\alpha\beta|C'|\alpha' \beta'}=\delta_{\alpha \alpha'} \sum_i \braket{\tilde{i}\tilde{i}|C|\beta' \beta}$, $\forall$ $\alpha, \beta, \alpha'$ and $\beta'$.

Next we prove that this condition in Eq.~\eqref{necessary} is sufficient for NCPTP-local passivity. Let us begin by considering the expression of $F$, given by
\begin{equation}
    F=\sum_{i,j} \Tr\left[C \left( U \ketbra{\tilde{j}}{\tilde{i}}U^{\dagger} \otimes \ket{\tilde{j}}\bra{\tilde{i}}\right) \right].
\end{equation}
Now, from the positive semidefiniteness of the state, $\ketbra{\Psi}{\Psi}$, and using inequality~\eqref{necessary}, we obtain
\begin{eqnarray}
   && F\ge 
\sum_{i,j} \Tr\left[C' \left( U \ketbra{\tilde{j}}{\tilde{i}}U^{\dagger} \otimes \ket{\tilde{j}}\bra{\tilde{i}}\right) \right] \nonumber \\
&=& \sum_{i,j,k_1,k_2} \Tr\left[C' \left( \ketbra{k_1}{k_2} \otimes \ket{\tilde{j}}\bra{\tilde{i}}\right) \right], \nonumber 
\end{eqnarray}
where we have defined $\ket{k_1}=U\ket{\tilde{i}}$, and $\ket{k_2}=U\ket{\tilde{j}}$.
The definition of $C'$ given in inequation~\eqref{necessary}, therefore suggests 
\begin{eqnarray}
    F \ge \delta_{k_1k_2} \sum_{i,j,k_1,k_2}\braket{k_1 \tilde{j}|C|k_2 \tilde{i}} &=& \sum_{s,i}\braket{ss|C|\tilde{i}\tilde{i}}. \nonumber \\
    &=& \Tr(\rho_{AB}H_{AB}).
\end{eqnarray}
The delta function gives $k_1=k_2$, which 
implies $U\ket{\tilde{i}}=U\ket{\tilde{j}}$, which in turn gives $\ket{\tilde{i}}=\ket{\tilde{j}}$, since $U$ is invertible.
This completes the proof that inequality~\eqref{necessary} is both necessary and sufficient for the pair, $\{H_{AB}, \rho_{EAB}\}$, to be NCPTP-local passive.
Additionally,  
we provide an independent necessary condition for NCPTP-local passivity using a different approach in the appendix.

\section{Independent Necessary conditions for NCPTP-local passivity}
\label{passive}
In this section, we derive a necessary condition for a state to be NCPTP-local passive for an arbitrary but fixed Hamiltonian, $H_{AB}$.
The maximum extractable energy using general LNCPTP operations from a state $\rho_{AB}$ which is a part of that of a bigger system, $\rho_{EAB}$, i.e., $\rho_{AB}=\Tr_E(\rho_{EAB})$, is
\begin{eqnarray*}
     && \Delta W^{NCP} (\rho_{AB},H_{AB})=  \text{Tr}\left[\rho_{AB}  H_{AB} \right] \nonumber \\ 
   &-& \min_{U_{EA}} \text{Tr}\big[\text{Tr}_E(U_{EA}\otimes {I}_B\rho_{EAB}U_{EA}^{\dagger}\otimes {I}_B) H_{AB} \big],
\end{eqnarray*}
where $U_{EA}$ is a unitary operation acting on the joint state of $E$ and $A$.
The second term on the right-hand side of the above equation can be written as $\min_{U_{EA}}\text{Tr}\big[(U_{EA}\otimes I_B\rho_{EAB}U_{EA}^{\dagger}\otimes I_B) I_E \otimes H_{AB} \big]$. So the condition of local passivity, i.e. $\Delta W^{NCP}_{\max} =0$, would imply $ \text{Tr}\big[(U_{EA}\otimes I_B\rho_{EAB}U_{EA}^{\dagger}\otimes I_B) I_E \otimes H_{AB} \big] \ge \text{Tr}\left[\rho_{EAB}  (I_E \otimes H_{AB}) \right]$ for all $U_{EA}$.
Therefore, our problem effectively reduces to finding the necessary condition for the relation $\text{Tr}\big[U_{a}\otimes I_b Y_{ab}U_{a}^{\dagger}\otimes I_b X_{ab} \big] \ge \text{Tr}\left[Y_{ab}  X_{ab} \right]$ to be true for all unitary operators $U_a$, where $Y_{ab}$ and $X_{ab}$ are general hermitian operators acting on the composite Hilbert space $\mathcal{H}_a\otimes\mathcal{H}_b$ and $I_b$ ($I_a$) is the identity operator acting on $\mathcal{H}_b$ ($\mathcal{H}_a$). To determine the condition, we consider that $\text{Tr}\big[U_{a}\otimes I_b Y_{ab}U_{a}^{\dagger}\otimes I_b X_{ab} \big] \ge \text{Tr}\left[Y_{ab}  X_{ab} \right]$ is true for an arbitrary $U_{a}$. We can expand the unitary operator $U_a$ in a power series
of an anti-Hermitian operator, say $M_a$, as
\begin{eqnarray}
    U_a &=& I_a+\frac{M_a}{1!}+\frac{M_a^2}{2!}+\frac{M_a^3}{3!}+\cdot\cdot\cdot. \label{3eq3}
\end{eqnarray}
Here it is considered that $\|M_a\| \ll 1$.
Decomposing $X_{ab}$ and $Y_{ab}$ most generally in a product basis and using Eq. \eqref{3eq3}, it can be shown that 
\begin{eqnarray}
\label{M}
    && \text{Tr}\big[U_{a}\otimes I_b Y_{ab}U_{a}^{\dagger}\otimes I_b X_{ab} \big] \nonumber \\
    &=& \text{Tr}\left[Y_{ab} , X_{ab} \right] 
    + \text{Tr} \left[(M_a \otimes I_b)[Y_{ab},X_{ab}]\right] + O(||M_a||^2).\nonumber
\end{eqnarray}
The minimum value of the left-hand side of the above equation can be equal to the first term on the right-hand side, i.e., $\text{Tr}\left[Y_{ab} , X_{ab} \right]$, only if the second term on the right-hand side, $\text{Tr} \left[(M_a \otimes I_b)[Y_{ab},X_{ab}]\right]$, becomes zero for all $M_a$.
We can express the second term as
\begin{eqnarray}
    \text{Tr} \left[(M_a \otimes I)[Y_{ab},X_{ab}]\right] &=& \text{Tr} \left(M_a \text{Tr}_b[Y_{ab},X_{ab}]\right).
\end{eqnarray}
The above term should be zero for all $\|M_a\|\ll 1$ which imply $\text{Tr}_b[Y_{ab},X_{ab}]=0$. Thus it offers the condition for $ \text{Tr}\big[U_{a}\otimes I_b Y_{ab}U_{a}^{\dagger}\otimes I_b X_{ab} \big] $ to be minimum at $U_a=I_a$.
Realizing $Y_{ab}$ as $\rho_{EAB}$ and $X_{ab}$ as $I_E \otimes H_{AB}$ the necessary condition for a state, $\rho_{AB}$, to be NCPTP-local passive reduces to 
\begin{equation}
\text{Tr}_B[\rho_{EAB},I_E \otimes H_{AB}]=0.\label{3eq4}
\end{equation}
It should be noted that if we restrict $\rho_{EAB}$ to have only classical correlation within the partition $E:AB$, Eq. \eqref{3eq4} would provide the condition for the state $\rho_{AB}$ to be CPTP-local passive.


Let us now analyze some additional constraints on the state, $\rho_{EAB}$, and Hamiltonian, $H_{AB}$, that should be satisfied by the state, $\rho_{AB}$, to be NCPTP-local passive, by restricting ourselves to a three-qubit Hilbert space for $EAB$. To find them, we will minimize the quantity $\text{Tr}\big[(U_{EA}\otimes I_B\rho_{EAB}U_{EA}^{\dagger}\otimes I_B) I_E \otimes H_{AB} \big]$ over a set of unitaries, $\{U_{EA}\}$, and impose the constraint that the minimum should be reached when $U_{EA}$ is identity. This considered set consists of block diagonal unitaries having two blocks of equal size.
The arbitrary $8\times 8$ and $4\times 4$ matrices, $\rho_{EAB}$ and $H_{AB}$, can be written as
\begin{eqnarray}
\label{xy}
    \rho_{EAB} = \sum_{i,j=1}^8a_{i,j}\kett{i}\braa{j}, \;\;\;
    \text{and}~~~
    H_{AB} = \sum_{i,j=1}^4b_{i,j}\kett{i}\braa{j},
\end{eqnarray}
where $a_{i,j}$ and $b_{i,j}$ are complex numbers and $\{\kett{i}\}_i$ is any arbitrary orthonormal basis.
The unitary operator, $U_{EA}$, is considered to be of the form
\begin{equation}
\label{U}
    U_{EA}=\left( {\begin{array}{cc}
       U_1 & 0 \\
        0 & U_2
        \end{array} } \right), 
\end{equation}
where $U_1,U_2 \in SU(2)$, and their explicit expressions are given by
\begin{equation}
    U_k=\left( {\begin{array}{cc}
        e^{i \theta_k}\cos{\phi_k} & e^{i\chi_k}\sin{\phi_k} \\
        -e^{-i \chi_k}\sin{\phi_k} & e^{-i\theta_k}\cos{\phi_k}
        \end{array} } \right).
        \label{unitary}
\end{equation}
Here, the index $k$ takes values $1$ and $2$ representing $U_1$ and $U_2$ respectively. $\theta_k\in [0,2\pi)$, $\chi_k\in [0,2\pi)$, and $\phi_k\in [0,\pi]$ are the free parameters of $U_k$.
We rechristen the quantity $\text{Tr}\big[(U_{EA}\otimes I_B\rho_{EAB}U_{EA}^{\dagger}\otimes I_B) I_E \otimes H_{AB} \big]$ as $W$ and optimize $W$ with respect to $\theta_k$, $\chi_k$, and $\phi_k$. We want $W$ to be minimum at $\theta_k=\chi_k=\phi_k=0$, which represents the identity operator.
The first derivatives of $W$ with respect to all the free parameters, i.e., $dW/d\theta_k , dW/d\chi_k $ and $ dW/d\phi_k $ do become zero at $\theta_k=\chi_k=\phi_k=0$, for $k=1,2$, under the condition, $\text{Tr}_B[\rho_{EAB},I_E \otimes H_{AB}]=0$.
Next we try to find the condition for the critical point, $\theta_k=\chi_k=\phi_k=0$, to be a minimum.
In this regard, we calculate the Hessian of $W$ with respect to $\theta_k,$ $\chi_k$ and $\phi_k$, at the point $\theta_k=\chi_k=\phi_k=0$ for $k=1,2$.
The Hessian matrix to be calculated in this case, has the following form
\begin{equation}
    \widetilde{H}=\left( {\begin{array}{cccccc}
      \frac{\partial^2W}{\partial\phi_1^2} & \frac{\partial^2W}{\partial\phi_1\partial\phi_2}  & \frac{\partial^2W}{\partial\phi_1\partial  \theta_1} & \frac{\partial^2W}{\partial\phi_1\partial  \chi_1} & \frac{\partial^2W}{\partial\phi_1\partial  \theta_2} & \frac{\partial^2W}{\partial\phi_1\partial  \chi_2} \\
       \frac{\partial^2W}{\partial\phi_2\partial\phi_1}  & \frac{\partial^2W}{\partial\phi_2^2} & \frac{\partial^2W}{\partial\phi_2\partial  \theta_1} & \frac{\partial^2W}{\partial\phi_2\partial  \chi_1} & \frac{\partial^2W}{\partial\phi_2\partial  \theta_2} & \frac{\partial^2W}{\partial\phi_2\partial  \chi_2} \\
       & & & & & \\
       \vdots & \vdots & \vdots & \vdots & \vdots & \vdots \\
        & & & & & \\
        \frac{\partial^2W}{\partial  \chi_2\partial\phi_1}  &\frac{\partial^2W}{\partial  \chi_2\partial\phi_2}  & \frac{\partial^2W}{\partial  \chi_2\partial  \theta_1} & \frac{\partial^2W}{\partial  \chi_2\partial  \chi_1} & \frac{\partial^2W}{\partial  \chi_2\partial  \theta_2}& \frac{\partial^2W}{\partial  \chi_2^2} \nonumber \\
        \end{array} } \right).
\end{equation}
By calculating the partial derivatives using the expressions of $\rho_{EAB}$ and $H_{AB}$ given in Eq.~\eqref{xy}, we get the Hessian matrix,
\begin{equation}
     \widetilde{H}=\left( {\begin{array}{cccccc}
       a  & 0  & b &c & 0 & 0\\
       0  & d & 0 & 0 & e & f \\
       b & 0 & h & 0 & 0 & 0 \\
       c & 0 & 0 & 0 & 0 & 0 \\
       0 & e & 0 & 0 & k & 0 \\
        0 & f  & 0 & 0 & 0 & 0 \\
        \end{array} } \right),
\end{equation}
where 
\begin{eqnarray}
    a&=&-2\sum_{\substack{i=1,2}}x_{i,i} \;\;\;\;
    b=\iota\sum_{\substack{i=1,2}}y_{i,i} \;\;\;\;
    c=\iota\sum_{\substack{i=1,2}}z_{i,i} \nonumber \\
    d&=&-2\sum_{\substack{i=5,6}}x_{i,i-4} \;\;\;\;
    e=\iota\sum_{\substack{i=5,6}}y_{i,i-4} \;\;\;\;
    f=\iota\sum_{\substack{i=5,6}}z_{i,i-4} \nonumber \\
    h&=&-4\sum_{\substack{i=1,2}}h_{i,i} \;\;\;\;
    k=-4\sum_{\substack{i=5,6}}h_{i,i-4}.
\end{eqnarray}
Here $\iota$ refers to $\sqrt{-1}$.
The explicit expressions of $x_{i,j}$,
$y_{i,j}$, $z_{i,j}$ and $h_{i,j}$ are given below
for completeness
 \begin{eqnarray}
    x_{i,j}&=&(a_{i,i}-a_{i+2,i+2}) (b_{j,j}-b_{j+2,j+2}) \nonumber \\
    &+& (a_{i,i+2}+a_{i+2,i}) (b_{j,j+2}+b_{j+2,j}) \nonumber \\
    &+& (a_{3-i,2+i}+a_{5-i,i})(b_{j,5-j}+b_{2+j,3-j}) \nonumber \\
    &+& (a_{3-i,i}-a_{5-i,2+i})(b_{j,3-j}-b_{2+j,5-j}),\nonumber 
 \end{eqnarray}
\begin{eqnarray}
    y_{i,j}&=&(a_{i,i}-a_{i+2,i+2}) (b_{j,j+2}-b_{j+2,j}) \nonumber \\
    &+& (a_{i,i+2}-a_{i+2,i}) (b_{j,j}-b_{j+2,j+2}) \nonumber \\
    &+& (a_{3-i,2+i}-a_{5-i,i})(b_{j,3-j}-b_{2+j,5-j}) \nonumber \\
    &+& (a_{3-i,i}-a_{5-i,2+i})(b_{j,5-j}-b_{2+j,3-j}),\nonumber 
 \end{eqnarray}
 \begin{eqnarray}
    z_{i,j}&=&(a_{i,i}-a_{i+2,i+2}) (b_{j,j+2}-b_{j+2,j}) \nonumber \\
    &-& (a_{i,i+2}-a_{i+2,i}) (b_{j,j}-b_{j+2,j+2}) \nonumber \\
    &-& (a_{3-i,2+i}-a_{5-i,i})(b_{j,3-j}-b_{2+j,5-j}) \nonumber \\
    &+& (a_{3-i,i}-a_{5-i,2+i})(b_{j,5-j}-b_{2+j,3-j}),\nonumber 
 \end{eqnarray}
 and
 \begin{eqnarray}
     h_{i,j}&=&(a_{i+2,i}b_{j,j+2}+a_{i,i+2}b_{j+2,j}) \nonumber \\
             &+& (a_{2+i,3-i}b_{3-j,2+j}+a_{i,5-i}b_{5-j,j}). \nonumber \\
 \end{eqnarray}
So the condition for $W$ to be minimum at the point, $  \theta_k=  \chi_k=\phi_k=0$, for $k=1,2$, is that, the Hessian matrix, $\widetilde{H}$, should be positive definite.
To conclude, the necessary conditions under which the two-qubit state state, $\rho_{AB}$, which is a part of a composite three-qubit state, $\rho_{EAB}$, is NCPTP-local passive with respect to the Hamiltonian, $H_{AB}$, are given by 
\begin{itemize}
    \item 
    $\text{Tr}_B[\rho_{EAB},I_E \otimes H_{AB}]=0$,
    \item
    $\widetilde{H}>0$.
\end{itemize}

\bibliography{battery_ncp}
\end{document}